\documentclass[%
 preprint,
 amsmath,
 amssymb,
 aps,
 nofootinbib,
 preprintnumbers
]{revtex4-1}
\usepackage{simplewick}
\usepackage{footnote}
\usepackage{graphicx}
\usepackage{dcolumn}
\usepackage{bm}
\usepackage{stmaryrd}
\usepackage{amsmath}
\usepackage{amssymb}
\usepackage{bbm}
\usepackage{amsfonts}
\usepackage{xcolor}
\usepackage{footmisc}
\usepackage[linktocpage=true,bookmarksnumbered=true,colorlinks=true,plainpages,linkcolor=blue,citecolor=red]{hyperref}

\newcommand{\centered}[1]{\begin{tabular}{l} #1 \end{tabular}}
\newcommand{\matrixbb}[4]{\left(\hspace{-5 pt}\begin{tabular}{ c c } ${#1}$ & ${#2}$ \\ ${#3}$ & ${#4}$ \end{tabular}\hspace{-5 pt}\right)}

\begin{document}
\preprint{MSUHEP-21-008}
\title{Spin-2 KK Mode Scattering in 
Models with a Massive Radion}


\author{R. Sekhar Chivukula$^{a}$}
\author{Dennis Foren$^{a}$}
\author{Kirtimaan A. Mohan$^{b}$}
\author{Dipan Sengupta$^{a}$}
\author{Elizabeth H. Simmons$^{a}$}

\affiliation{$^{a}$ Department of Physics and Astronomy, 9500 Gilman Drive,
 University of California, San Diego }
 \affiliation{$^{b}$ Department of Physics and Astronomy, 567 Wilson Road, Michigan State University, East Lansing}
\email{rschivukula@physics.ucsd.edu}
\email{dennisforen@gmail.com}
\email{kamohan@msu.edu}
\email{disengupta@physics.ucsd.edu}
\email{ehsimmons@ucsd.edu}

\date{\today}

\begin{abstract}
We calculate tree-level scattering amplitudes of massive spin-2 KK particles in models of stabilized compact extra-dimensional theories.  Naively introducing a mass for the radion in an extra-dimensional model without accounting for the dynamics responsible for stabilizing the extra dimension upsets the cancellations relating the masses and couplings of the spin-2 modes, resulting in KK scattering amplitudes which grow like $E^{4}$ instead of $E^{2}$. We therefore investigate scattering of the Kaluza-Klein states in theories incorporating the Goldberger-Wise mechanism to stabilize the size of the extra dimension. We demonstrate that the cancellations occur only when one includes not only the massive radion, but also the massive spin-0 modes arising from the Goldberger-Wise scalar. We compute the revised sum rules which are satisfied in a stabilized model to ensure a consistent high-energy scattering amplitude. We introduce a simple model of a stabilized extra dimension which is a small deformation of a flat (toroidal) five-dimensional model, and demonstrate the cancellations in computations performed to leading nontrivial order in the deformation. These results are the first complete KK scattering computation in an extra-dimensional model with a stabilized extra dimension, with implications for the theory and phenomenology of these models.
\end{abstract}

\maketitle

\tableofcontents

\section{Introduction}

Examining the possibility of extra spatial dimensions has a long history  \cite{Kaluza:1921tu,Klein:1926tv,Appelquist:1987nr}, and has been of particular interest for constructing possible solutions to the hierarchy problem \cite{Antoniadis:1990ew,ArkaniHamed:1998rs,Appelquist:2000nn,Randall:1999ee,Randall:1999vf}. To be consistent with observation, four of the  dimensions must correspond to ordinary spacetime, and viable backgrounds must include (at least approximate) Poincar\'{e} invariance in these extended directions. Each of the fields on the full spacetime can then be decomposed into an infinite  Kaluza-Klein (KK) tower of four-dimensional states of different masses, with the mass-scale of the tower of states (typically) set by the size of the transverse extra dimensions. The effective four-dimensional theory then consists of interacting KK modes. The higher-dimensional graviton field, in particular, gives rise to many 4D states: the ordinary four-dimensional graviton and an infinite KK tower of massive spin-2 states, and potentially 4-vector and scalar states as well. In the five-dimensional theories which we consider here, we impose an orbifold symmetry on the internal space and use a gauge freedom to eliminate any gravitational 4-vector states from the higher-dimensional metric; a scalar portion of the metric remains unfixed, however, giving rise to a (five-dimensional) radion field.

In recent work we examined the scattering amplitudes of the massive spin-2 modes \cite{Chivukula:2019rij,Chivukula:2019zkt,Chivukula:2020hvi,Foren:2020egq} in compactified five-dimensional theories\footnote{See also \cite{Bonifacio:2019ioc}.}. We demonstrated that an intricate set of cancelling contributions from the exchange of KK modes of different levels enables these scattering amplitudes to avoid the bad high-energy behavior that typically plagues models with interacting massive spin-2 particles \cite{ArkaniHamed:2002sp,ArkaniHamed:2003vb,Hinterbichler:2011tt}. In particular, the elastic scattering amplitudes of interacting helicity-0 massive spin-2 states have individual contributions which grow as fast as $E^{10}$. We have shown that a set of four independent sum rules relating the masses and couplings of the various KK modes reduces these amplitudes to the $E^2$ growth which would be expected in a consistent theory including four-dimensional gravity.

In models where gravity propagates in the extra spatial dimension, the overall size of the extra dimension is undetermined. This size can be associated with the background value of the radion field. Therefore, after compactification, perturbations corresponding to fluctuations in the size of the transverse extra dimension manifest as a massless four-dimensional radion scalar field (for a review, see \cite{Rattazzi:2003ea}).\footnote{In extra-dimensional models dual to conformal field theories \cite{Maldacena:1997re,Witten:1998qj,Aharony:1999ti}, the radion is dual to the dilaton expected after the spontaneous breaking of scale symmetry.} Note there are two very different classes of problems where a 5D compactified theory of gravity might be relevant: as an extra-dimensional model of spacetime \cite{Randall:1999ee,Randall:1999vf}, or as the ``dual" of a TeV-scale strongly-coupled 4D theory \cite{Maldacena:1997re,Witten:1998qj,Aharony:1999ti}. In either case, a massless radion is phenomenologically forbidden. In the case of ordinary gravity, a massless radion contributes to long-range gravitational interactions as a Brans-Dicke scalar \cite{Brans:1961sx} and thus generates significant deviations from general relativistic predictions already confirmed by experiments, including the gravitational deflection of light.  In the case of a warped model dual to a strongly-coupled conformal theory at the TeV scale (and in which one considers adjusting parameters to decouple the graviton), the massless radion has TeV-scale couplings and runs into conflict with astrophysical constraints \cite{Abu-Ajamieh:2017khi}. Consequently, any phenomenologically relevant model must include additional dynamics to stabilize the size of the extra dimension and give mass to the radion.

Our previous computations of the amplitudes describing massive spin-2 KK mode elastic scattering were performed in unstabilized five-dimensional theories and included contributions to the scattering amplitudes from a massless radion. Indeed, contributions from a massless radion were crucial in these calculations to prevent the scattering amplitudes from growing like $E^6$ or $E^4$. Simply introducing a radion mass by hand (and assuming its couplings are otherwise unchanged) breaks the underlying higher-dimensional diffeomorphism invariance, and results in scattering amplitudes which diverge like $E^4$ \cite{Chivukula:2019rij,Chivukula:2019zkt,Chivukula:2020hvi,Schwartz:2003vj}. In a theory with a properly stabilized extra dimension, however, divergences beyond $E^{2}$ growth should not occur.

Goldberger and Wise (GW) \cite{Goldberger:1999uk,Goldberger:1999un} introduced a simple dynamical mechanism for stabilizing the size of an extra dimension.  In the GW model, one adds a bulk scalar field whose dynamics are chosen so that the vacuum of the system has a nonconstant profile for the scalar field in the extra dimension. On energetic grounds, Goldberger and Wise showed that competition between the contributions from the scalar kinetic energy and potential energy ({\it e.g.} from a mass term) imply that there is a preferred size for the extra dimension. Since the radion field can be identified with fluctuations in the size of the extra dimension, the existence of a preferential extra-dimensional distance scale in the GW model directly generates a radion mass \cite{Tanaka:2000er,Csaki:2000zn}.

In this paper, we compute the tree-level elastic 2-to-2 scattering amplitude of helicity-0, KK level-$n$, massive spin-2 KK modes in a GW model with a stabilized extra dimension; we identify generalized sum rules controlling the growth of the scattering amplitude; and we show that the cancellations between the different contributions to the scattering amplitude persist once one includes the new contributions to massive spin-2 KK mode scattering from the GW sector. Our calculation hinges on properly identifying the propagating degrees of freedom in the scalar part of the five-dimensional metric and bulk scalar sectors, after accounting for diffeomorphism invariance \cite{Csaki:2000zn,Kofman:2004tk,Boos:2005dc,George:2011sw,Chivukula:2020tbd}. 
 
Radion mass generation in the GW model is similar, in principle, to what happens in a spontaneously broken gauge theory. In that case, the longitudinal components of the massive gauge bosons (which correspond to the broken symmetries) mix with derivatives of the Goldstone bosons through terms proportional to the vacuum expectation value responsible for the gauge symmetry breaking. Gauge-fixing eliminates one combination of these fields, and the physical degrees of freedom which remain are massive spin-1 fields. Accounting for additional contributions from the symmetry breaking sector ({\it e.g.} the Higgs boson) is thus crucial to understanding why the scattering amplitudes of the massive spin-1 bosons do not diverge at high energies.
 
In the case of the GW mechanism, scalar particles seemingly originate from either of two unrelated scalar towers: the scalar perturbation of the five-dimensional metric (which generates the radion mode) and the fluctuation of the bulk scalar field configuration. However, as emphasized by a particular gauge choice that fixes a certain linear combination of these fields, the metric perturbation actually mixes with the {\it derivative} of the bulk scalar fluctuation, ultimately yielding a single physical 4D scalar tower. We will refer to the lightest mode in this tower as the ``radion" and the higher modes as ``GW scalars".\footnote{The gauge-fixing described is the analog of the choice of coordinates that reduces the physical degrees of freedom in scalar metric perturbations down to a single massless 4D radion field in the unstabilized model \cite{Charmousis:1999rg}.} Properly identifying and normalizing these modes \cite{Kofman:2004tk,Boos:2005dc,George:2011sw,Chivukula:2020tbd} is crucial in order to compute their couplings to the massive spin-2 modes. We explicitly compute the revised sum rules that are satisfied in a stabilized GW model to ensure a consistent high-energy scattering amplitude. 
 
Having identified the generalized sum rules which must be satisfied, we
next introduce a simple model of a stabilized extra dimension in which we can explicitly demonstrate these cancellations. To do so, we cannot treat the background spacetime geometry as fixed: ignoring the ``backreaction" of the background scalar field on the geometry leaves the radion massless  \cite{Tanaka:2000er,Csaki:2000zn}. DeWolfe, Freedman, Gubser and Karch (DFGK) \cite{DeWolfe:1999cp} determined an entire class of exact solutions wherein a bulk scalar field and standard five-dimensional gravity successfully and self-consistently exhibit the GW mechanism.  In this paper, we introduce a simple limit of the DFGK model, which we dub the  ``flat stabilized model". 
As we will detail, the flat stabilized model results in a geometry which is a small deformation of a five-dimensional orbifolded toroidally compactified space; in this geometry, all quantities pertaining to the spin-2 and scalar KK modes can be computed analytically to the required order in the deformation parameter. As the deformation parameter goes to zero, the radion becomes massless (justifying our labeling it as a radion) while the GW scalars remain massive. The flat stabilized model provides a simple explicit realization of the GW mechanism and allows us to illustrate how the sum rules maintain their validity as the stabilized geometry becomes relevant.

These results are the first complete scattering computation in an extra-dimensional model with a stabilized extra dimension. The generalization of our results to the case of a stabilized warped model, a description of phenomenological implications, and a more detailed derivation and analysis of the bulk scalar and scalar metric sector will be presented in \cite{Chivukula:2020tbd}.

In section \ref{sec:GW-mechanism-2}, we describe the Goldberger-Wise mechanism, and also the spin-2 and spin-0 KK modes that result. Section \ref{sec:spin-0-modes}, in particular, describes the scalar modes which result after mixing between the bulk scalar and scalar metric sectors of the theory. In section \ref{sec:KK-scattering-3}, we describe the KK mode couplings and compute the elastic scattering amplitude of helicity-0, KK level-$n$, massive spin-2 KK modes. In section \ref{subsec:sum-rules}, we identify the generalized sum rules: the combinations of masses and couplings which must vanish if the scattering amplitude is to grow no faster than ${\mathcal O}(s)$. We describe how we can prove three out of four of these sum rules analytically for any model incorporating the GW mechanism. Section \ref{sec:flat-stabilized-model} introduces the flat stabilized model; summarizes the KK mode wavefunctions, masses, and couplings that result; and demonstrates that the sum rules are explicitly satisfied in this model to second order in the deformation parameter. Our conclusions are presented in section \ref{sec:Conclusions}.
\section{KK Modes and the GW Mechanism}

\label{sec:GW-mechanism-2}

In this section, we review the dynamics of the Goldberger-Wise (GW) mechanism \cite{Goldberger:1999uk,Goldberger:1999un} for stabilizing the geometry of a five-dimensional theory and generating the mass of the radion. We set our notation for the metric, specify the interactions of the model, describe the origin of mixing between the bulk scalar and scalar metric modes, and establish the mode expansion for the physical four-dimensional modes in the gravitational and scalar sectors \cite{Kofman:2004tk,Boos:2005dc,George:2011sw,Chivukula:2020tbd}.

\subsection{Notation, Dynamics, and Gauge-Fixing}

\label{subsec:model}

The metric for a space having a warped extra dimension is $ds^2 = G_{MN} dx^M dx^N$ with coordinates $x^{M}=(x^{\mu},y)$, where $x^{\mu}$ parameterizes the usual 4D spacetime. We take the coordinate of the extra dimension to be  $y \equiv \varphi r_{c} \in [-\pi r_{c},+\pi r_{c}]$, and impose an orbifold symmetry $y\leftrightarrow-y$. Brane-localized dynamics are present at the orbifold fixed points  $\varphi\in\{0,\,\pi\}$, and the metric and brane-projected metric equal:
\begin{align}
    [G_{MN}] = \matrixbb{w\,g_{\mu\nu}}{0}{0}{-v^{2}}\hspace{35 pt} [\overline{G}_{MN}] = \matrixbb{w\,g_{\mu\nu}}{0}{0}{0}
\end{align}
respectively, where
\begin{align}
     w = e^{-2[A(y)+\hat{u}(x,y)]}~,\hspace{35 pt}&v = 1 + 2\hat{u}(x,y)~,\\  g_{\mu\nu} = \eta_{\mu\nu} + \kappa\, \hat{h}_{\mu\nu}(x,y)~,\hspace{35 pt}&\hat{u} = \dfrac{ e^{2A(y)}}{2\sqrt{6}} \kappa\,\hat{r}(x,y)~. \label{eq:uhat}
\end{align}
The function $A(y)$ is the warp factor for the background geometry and is determined by solving the Einstein field equations, while $\hat{h}_{\mu\nu}(x,y)$ and $\hat{r}(x,y)$ represent tensor and scalar metric perturbations around this geometry.\footnote{The form in which the radion perturbations are introduced eliminates kinetic mixing between the tensor and scalar metric perturbations \cite{Charmousis:1999rg}.}$^{,}$\footnote{The definition of the field $\hat{u}$ used here differs slightly from the corresponding definition used in \cite{Chivukula:2020hvi,Foren:2020egq} for the unstabilized Randall-Sundrum model: the definition used here is more convenient for analyzing the mixed bulk scalar and scalar metric sectors of the stabilized theory.} We will denote inverse matrices  with tildes, e.g. $[\tilde{G}^{MN}] = \text{Diag}(\tilde{g}^{\mu\nu}/w,-1/v^{2})$. The Lorentz metric is in the mostly-minus convention: $\eta_{\mu\nu}\equiv \text{Diag}(+1,-1,-1,-1)$.

The general Lagrangian for a GW model consists of three parts
\begin{align}
    \mathcal{L}_{\text{5D}} \equiv \mathcal{L}_{\text{EH}} + \mathcal{L}_{\Phi\Phi}+
    \mathcal{L}_{\text{pot}}~,
\end{align}
which we now define.
First, the theory includes the gravitational five-dimensional Einstein-Hilbert Lagrangian
\begin{align}
    \mathcal{L}_{\text{EH}} &\equiv -\dfrac{2}{\kappa^{2}} \sqrt{G}\, R~,
\end{align}
which provides the dynamics of the metric, where $\kappa^2=4/M_{\text{Pl,5D}}^{3}$ defines the five-dimensional gravitational constant. $M_{\text{Pl,5D}}$ is the 5D Planck scale.
Next, a bulk real scalar field $\hat{\Phi}(x,y)$ is included via a standard bulk scalar kinetic Lagrangian
\begin{align}
   \mathcal{L}_{\Phi\Phi} &\equiv  \sqrt{G}\bigg[\dfrac{1}{2}\,\tilde{G}^{MN}\,(\partial_{M}\hat{\Phi})\,(\partial_{N}\hat{\Phi})\bigg]~.
   \label{eq:scalarbulkLag}
\end{align}
Finally, there are potential energy terms for the bulk scalar field both in the bulk and at the branes,
\begin{align}
     \mathcal{L}_{\text{pot}}&\equiv -\,\dfrac{4}{\kappa^2}\bigg[ \sqrt{G} \,V[\hat{\Phi}] + \sqrt{\overline{G}}\,V_1[\hat{\Phi}]\delta(\varphi)
     +\sqrt{\overline{G}}\,V_2[\hat{\Phi}]\delta(\varphi-\pi)\bigg]~.
\end{align}
The index $i\in\{1,2\}$ will generally indicate the $\varphi = 0$ and $\varphi = \pi$ branes respectively.

We will show that the lightest state in the scalar tower generated by $\hat{r}(x,y)$ should be identified with the radion. In order to stabilize the geometry and give mass to the radion, the brane potential terms $V_{1,2}$ cause the field $\hat{\Phi}(x,y)$ to gain a nonconstant background value.\footnote{In the unstabilized limit, in which the bulk scalar field has no $y$-dependent background value and the radion is massless, the potential energy terms include the bulk and brane cosmological constants of the RS model \cite{Randall:1999ee,Randall:1999vf}.}
This scalar background will be some $y$-dependent function $\phi_{0}(y)$, relative to which we measure an $(x,y)$-dependent scalar fluctuation $\hat{f}(x,y)$. Explicitly, we define
\begin{align}
    \hat{\Phi}(x,y) \equiv \dfrac{1}{\kappa} \hat{\phi}(x,y) \equiv \dfrac{1}{\kappa}\bigg[\phi_{0}(y) + \hat{f}(x,y)\bigg]~,
\end{align}
where we have included a factor of $\kappa$ so that $\phi_{0}$ and $\hat{f}$ are dimensionless. This also neatly absorbs factors of $\kappa$ that will emerge from the gravitational sector of the calculation.

In the presence of a $y$-dependent scalar background $\phi_0(y)$, the scalar kinetic Lagrangian of Eq. \eqref{eq:scalarbulkLag} induces mixing between bulk scalar fluctuations and the scalar sector of the metric as described in the interaction
\begin{equation}
    {\mathcal L}_{\Phi\Phi} \supset  \dfrac{\sqrt{G}}{2}\tilde{G}^{55}(\partial_y \hat{\Phi})(\partial_{y}\hat{\Phi})\hspace{15 pt} \Rightarrow \hspace{15 pt} \hat{r} (\partial_{y} \phi_{0}) (\partial_{y} \hat{f})~.
\end{equation}
The scalar metric modes also have three-point couplings to the massive spin-2 KK modes. In the unstabilized RS model, those three-point couplings enable exchanges of particles from the scalar metric sector  that are crucial to cancelling the otherwise bad high-energy behavior of massive spin-2 KK mode scattering amplitudes \cite{Chivukula:2019rij,Chivukula:2019zkt,Chivukula:2020hvi,Foren:2020egq}. The mixing here implies that in an extra-dimensional model stabilized via the GW mechanism, the bulk scalar inherits three-point couplings to the massive spin-2 modes through mixing and thereby generates important contributions to their scattering amplitudes.

Analyzing the dynamics of the theory specified by ${\mathcal L}_{\text{5D}}$ therefore reduces to determining the background metric function $A(y)$ and scalar field configuration $\phi_0(y)$ consistent with the equations of motion; writing the theory in terms of the fluctuation fields $\hat{h}$, $\hat{r}$, and $\hat{f}$; expanding these into KK modes; and ``diagonalizing" the kinetic terms of these 4D fields. The analysis of the spin-2 fluctuations is straightforward and, as we describe below, proceeds in parallel with the  analyses used previously in \cite{Chivukula:2020hvi,Foren:2020egq}.

The analysis of the mixed bulk scalar and scalar metric sector, however, is more complicated. 
Due to the diffeomorphism invariance of the action \cite{Csaki:2000zn,Kofman:2004tk,Boos:2005dc,George:2011sw,Chivukula:2020tbd}, the bulk scalar $\hat{f}(x,y)$ and scalar metric $\hat{r}(x,y)$ fluctuation fields are not truly independent dynamical degrees of freedom -- only a single linear combination of the bulk scalar and scalar metric modes remains in the physical spectrum. In particular, we choose a gauge such that $\hat{f}$ and $(\partial_{\varphi}\hat{r})$ are proportional (recall that $y\equiv r_c \varphi$):
\begin{align}
    (\partial_{\varphi}\phi_{0})\,\hat{f}(x,y) \equiv \sqrt{6}\, e^{2A}\,\big[\kappa\,(\partial_{\varphi}\hat{r})\big]~, \label{SRS1GaugeCondition}
\end{align}
and will compute the spectrum and couplings in this gauge.\footnote{Note that in the unstabilized limit, in which $\phi_{0}$ becomes constant in the extra dimension, this gauge choice implies $(\partial_{\varphi} \hat{r})$ vanishes and therefore the field $\hat{r}$ gives rise to the single (massless) 4D radion field field which is present in the unstabilized model \cite{Charmousis:1999rg}.}

After applying this gauge, extremizing the Lagrangian implies the following background field equations \cite{DeWolfe:1999cp,Tanaka:2000er,Csaki:2000zn},
\begin{align}
    (\partial_{\varphi}^{2}A) &= \dfrac{1}{12}(\partial_{\varphi}\phi_{0})^{2} + 2(\partial_{\varphi}A)\Big[\delta(\varphi)-\delta(\varphi-\pi)\Big]\\
    (\partial_{\varphi}^{2}\phi_{0}) &= 4\dfrac{dV}{d\hat{\phi}}\bigg|_{\hat{\phi}=\phi_{0}} + 4(\partial_{\varphi}A)(\partial_{\varphi}\phi_{0}) + 2(\partial_{\varphi}\phi_{0}) \Big[\delta(\varphi) - \delta(\varphi-\pi)\Big]\\
    V[\phi_{0}] &= \dfrac{1}{8}(\partial_{\varphi}\phi_{0})^{2} - 6 (\partial_{\varphi} A)^{2}
\end{align}
and the following jump conditions,
\begin{align}
    V_{1,2}[\phi_{0}] = \pm 6(\partial_{\varphi}A) \hspace{35 pt} \dfrac{dV_{1,2}}{d\hat{\phi}} \bigg|_{\hat{\phi}=\phi_{0}}=\pm \dfrac{1}{2} (\partial_{\varphi}\phi_{0})
\end{align}
at the $\varphi=0$ and $\varphi=\pi$ branes respectively. These constrain the scalar field background configuration $\phi_{0}$; the scalar potentials $V$, $V_{1}$, and $V_{2}$; and the warp factor $A$.

In the next section we outline the results of the analysis needed for our computations here. More details will be given in a subsequent publication \cite{Chivukula:2020tbd}.

\subsection{The Kaluza Klein Modes}

We now determine the Kaluza-Klein spin-0 and spin-2 mode equations and orthonormality conditions. Here we present the results of our computations, which build on the results of \cite{Kofman:2004tk,Boos:2005dc,George:2011sw}, and discuss how they are related to our previous computations \cite{Chivukula:2019rij,Chivukula:2019zkt,Chivukula:2020hvi,Foren:2020egq}. For details of the derivations, see \cite{Chivukula:2020tbd}.

\subsubsection{Spin-2}

The spin-2 KK mode decomposition proceeds in the usual way and as described in \cite{Chivukula:2020hvi,Foren:2020egq}, except with the Randall-Sundrum warp factor $\varepsilon = e^{kr_{c}|\varphi|}$ used there replaced here with the more generic $\varepsilon = e^{A(\varphi)}$. In particular, 
the spin-2 modes can be decomposed via
\begin{align}
    \hat{h}_{\mu\nu}(x,y=r_c\varphi) = \dfrac{1}{\sqrt{\pi r_{c}}} \sum_{n=0}^{+\infty} \hat{h}^{(n)}_{\mu\nu}(x)\, \psi_{n}(\varphi)~,
    \label{eq:spin2mode-equation}
\end{align}
where the spin-2 particle described by $\hat{h}^{(n)}_{\mu\nu}(x)$ has mass $m_{n} = \mu_{n}/r_{c}$ and $\hat{h}^{(0)}_{\mu\nu}(x)$ describes the massless graviton. The wavefunctions $\psi_n(\varphi)$ satisfy the mode equation
\begin{align}
    \partial_{\varphi}\bigg[e^{-4A}(\partial_{\varphi}\psi_{n})\bigg] = -\mu_{n}^{2} e^{-2A} \psi_{n}~,
    \label{eq:spin2-SLeqn}
\end{align}
subject to the boundary conditions $(\partial_{\varphi} \psi_{n}) = 0$ at $\varphi \in \{0,\pi\}$. The Sturm-Liouville nature of the problem defined by this mode equation and boundary conditions ensures that the modes are
orthogonal and complete, and we obtain canonical kinetic energy terms for the 4D modes if the spin-2 wavefunctions are normalized according to
\begin{align}
    \dfrac{1}{\pi}\int_{-\pi}^{+\pi} d\varphi\hspace{10 pt} e^{-2A}\psi_{m}\psi_{n} &= \delta_{m,n}
    \label{eq:spin2-mode-norm}
\end{align}

\subsubsection{Spin-0}

\label{sec:spin-0-modes}

Using the gauge condition in Eq. \eqref{SRS1GaugeCondition}, we can eliminate the bulk scalar fluctuation field $\hat{f}(x,y)$ in favor of the scalar metric fluctuation field  $\hat{r}(x,y)$. 
We decompose the $\hat{r}(x,y)$ field  using the mode expansion\footnote{In our analysis of the unstabilized RS model \cite{Chivukula:2019rij, Chivukula:2019zkt,Chivukula:2020hvi,Foren:2020egq}, the corresponding field contained (after gauge-fixing \cite{Charmousis:1999rg}) only the massless four-dimensional radion field with a constant radion wavefunction. For convenience, in that work we chose $\hat{u}$ such that the radion wavefunction equalled the graviton wavefunction $\psi_0$.
The choice of the field $\hat{u}$  in Eq. \eqref{eq:uhat} differs from the one we made previously, and the mode $\gamma_0$ here reduces, in limit of an unstabilized RS model, to $e^{+kr_c\pi}\psi_0$.}
\begin{align}
    \hat{r}(x,y=r_c \varphi) = \dfrac{1}{\sqrt{\pi r_{c}}} \sum_{i=0}^{+\infty} \hat{r}^{(i)}(x)\, \gamma_{i}(\varphi)~.
    \label{eq:spin0-mode-expansion}
\end{align}
Each extra-dimensional spin-0 wavefunction $\gamma_{i}$ solves the Sturm-Liouville-{\it like} equation \cite{Kofman:2004tk,Boos:2005dc,George:2011sw,Chivukula:2020tbd}
\begin{align}
    \partial_{\varphi} \bigg[\dfrac{e^{2A}}{(\partial_{\varphi}\phi_{0})^{2}} (\partial_{\varphi}\gamma_{n})\bigg] - \dfrac{e^{2A}}{6} \gamma_{n} = - \mu_{(n)}^{2}\dfrac{e^{4A}}{(\partial_{\varphi}\phi_{0})^{2}} \gamma_{n}~,
    \label{eq:spin0mode-equation}
\end{align}
where each $\gamma_{i}$ corresponds to a 4D scalar state $\hat{r}^{(i)}$ with mass $m_{(i)} \equiv \mu_{(i)}/r_{c}$, subject to the mixed boundary conditions 
\begin{align}
    \bigg\{\bigg[
    2\,\dfrac{d^2V_1}{d\hat{\phi}^2}-\dfrac{(\partial^2_\varphi \hat{\phi})}{(\partial_\varphi \hat{\phi})} \bigg]
    (\partial_\varphi \gamma_n) + \mu^2_{(n)} e^{2A}\gamma_n\bigg\}\bigg|_{\hat{\phi}=\phi_0,\,\varphi=0^{+}} & =0~, \nonumber \\
     \bigg\{\bigg[
    2 \, \dfrac{d^2V_2}{d\hat{\phi}^2}+\dfrac{(\partial^2_\varphi \hat{\phi})}{(\partial_\varphi \hat{\phi})} \bigg]
    (\partial_\varphi \gamma_n) - \mu^2_{(n)} e^{2A}\gamma_n\bigg\}\bigg|_{\hat{\phi}=\phi_0,\,\varphi=\pi^{-}} & =0~.
    \label{eq:spin0-bcs}
\end{align}

By including contributions from both the bulk scalar and scalar metric sectors of the theory, using the equations of motion, attending to boundary contributions, and imposing the normalization condition
\begin{equation}
    \dfrac{6}{\pi} \int_{-\pi}^{\pi} ~d\varphi\bigg[ \dfrac{e^{2A}}{(\partial_{\varphi}\phi_{0})^{2}}\, (\partial_{\varphi}\gamma_{m}) (\partial_{\varphi}\gamma_{n}) + \dfrac{e^{2A}}{6}\gamma_{m}\gamma_{n} \bigg] =\delta_{mn}~,
    \label{eq:scalarnorm1}
\end{equation}
the quadratic scalar mode Lagrangian can be written as \cite{Kofman:2004tk,Boos:2005dc,George:2011sw,Chivukula:2020tbd}
\begin{equation}
    \mathcal{L}_{\gamma\gamma} = \sum_{n=0}^{+\infty}\bigg\{\dfrac{1}{2} \Big(\partial_{\mu}\hat{r}^{(n)}(x)\Big)\Big(\partial^{\mu}\hat{r}^{(n)}(x)\Big) - \dfrac{\mu_{(n)}^2}{2\,r_{c}^{2}} \, \hat{r}^{(n)}(x)\,\hat{r}^{(n)}(x)\bigg\}~.
    \label{scal-lag}
\end{equation}
The second term in the integrand of Eq. \eqref{eq:scalarnorm1} arises from the scalar metric sector of the theory, while the unconventional first term arises from the bulk scalar sector via the gauge condition Eq. \eqref{SRS1GaugeCondition} used to eliminate $\hat{f}$ in favor of $\partial_\varphi \hat{r}$.
While these orthogonality and normalization conditions are consistent \cite{Kofman:2004tk,Boos:2005dc,George:2011sw} with the mode differential equation  and boundary conditions of Eqs. \eqref{eq:spin0mode-equation} and \eqref{eq:spin0-bcs}, they are {\it not} the normalization conditions required if we were to interpret them as defining a Sturm-Liouville problem. Therefore, while completeness of the mode expansion in \eqref{eq:spin0-mode-expansion} is physically reasonable, it is not mathematically guaranteed. The sum rules we consider in the next section apply so long as the mode expansion
in Eq. \eqref{eq:spin0-mode-expansion} is valid.

The mathematical uncertainties of the scalar sector are eliminated, however,  if we work in the ``stiff-wall" limit \cite{Goldberger:1999uk,Goldberger:1999un}, such that 
\begin{equation}
    \dfrac{d^2 V_{1}}{d\hat{\phi}^{\,2}}\bigg|_{\varphi=0^{+}}\hspace{5 pt}\text{ and }\hspace{5 pt} \dfrac{d^2 V_{2}}{d\hat{\phi}^{\,2}}\bigg|_{\varphi=\pi^{-}}\hspace{15 pt} \to +\infty~.
    \label{eq:stiff-wall}
\end{equation}
In this case, the boundary conditions of Eq. \eqref{eq:spin0-bcs} reduce to the Neumann conditions
$(\partial_{\varphi}\gamma_{n}) = 0$ at $\varphi \in \{0,\pi\}$. Using Eqs. \eqref{eq:spin0mode-equation} and \eqref{eq:spin0-bcs} as well as integrating by parts, we find that the orthonormality conditions can be rewritten as \cite{Kofman:2004tk,Boos:2005dc,George:2011sw}
\begin{align}
    \dfrac{6}{\pi}\,\mu_{(n)}^{2} \int_{-\pi}^{+\pi} d\varphi\hspace{10 pt} \dfrac{e^{4A}}{(\partial_{\varphi}\phi_{0})^{2}} \gamma_{m}\gamma_{n} &= \delta_{m,n}~.
    \label{eq:scalarnorm2}
\end{align}
in the stiff wall limit. Hence, so long as $\mu^2_{(n)} >0$, the inner product has the proper form to interpret the spin-0 mode equation Eq. \eqref{eq:spin0mode-equation} as a Sturm-Liouville equation with weight function $e^{4A}/(\partial_\varphi \phi_0)^2$ -- and completeness in the stiff wall limit is assured. The explicit model we analyze in section \ref{sec:flat-stabilized-model} will employ the stiff wall limit in order to simplify our results, and we will confirm that the sum rules we derive are explicitly satisfied in that case.

Because both forms of the orthonormality conditions, Eqs. \eqref{eq:scalarnorm1} and \eqref{eq:scalarnorm2}, contain derivatives of the scalar background $(\partial_\varphi \phi_0)^2$ in denominators, taking the limit of $\phi_0$ going to a constant -- and thereby connecting with the computations in the unstabilized model \cite{Chivukula:2019rij,Chivukula:2019zkt,Chivukula:2020hvi,Foren:2020egq}-- is not straightforward. In the unstabilized limit, the mixing between the bulk scalar and scalar metric sectors vanishes -- and, physically, the theory reduces to one with a massless radion and a tower of massive scalar KK modes which couple conventionally (through their energy-momentum tensor) with the gravitational sector. As we will demonstrate when we consider an explicit model, in the limit that $(\partial_\varphi \phi_{0}) \to 0$, the wavefunction $\gamma_0$ remains finite while $\mu^2_{(0)}$ vanishes. We therefore associate the lightest scalar KK mode $\hat{r}^{(0)}$ with the radion. Conversely, in this same limit, the wavefunctions of the higher scalar modes ($\gamma_n$ for $n >0$) vanish  -- indicating that their mixing with the scalar metric sector vanishes\footnote{Note that any couplings arising from the interactions of $\hat{\Phi}$, encoded in the fluctuations $\hat{f}$, remain finite because of the form of Eq. \eqref{SRS1GaugeCondition}. This also implies that the stiff wall Neumann boundary conditions on $\gamma_n$ are  Dirichlet  boundary  conditions  for  the  GW scalars, and therefore the GW scalar tower has no massless mode in the limit $(\partial_\varphi \phi_{0}) \to 0$.} -- while their masses $\mu^2_{(n)}$ remain finite. We will therefore distinguish the higher states $\hat{r}^{(n)}$ as ``GW scalars."

In the next section, we consider the sum rules that must be satisfied in order to obtain consistent behavior for the high-energy massive spin-2 scattering amplitudes.
\section{KK Scattering Amplitudes and Sum Rules}

\label{sec:KK-scattering-3}

In this section, we compute the scattering amplitude for elastic scattering of helicity-0, KK level-$n$, massive spin-2 KK bosons in a general GW model, specifically for the process $n,n \to n,n$. We demonstrate that requiring this amplitude to grow no faster than ${\cal O}(s)$ imposes sum rules that relate the masses and couplings of the spin-2 and spin-0 KK modes. These sum rules generalize those presented in \cite{Chivukula:2019zkt,Bonifacio:2019ioc}, in order to apply to models incorporating the GW mechanism, which stabilizes the size of the extra dimension and makes the radion massive. Crucially, the $\mathcal{O}(s^{3})$ and $\mathcal{O}(s^{2})$ growth of this amplitude each only cancel once one includes contributions from the GW scalars and their masses.

\begin{figure}
    \centering
    \includegraphics[width=0.8\textwidth]{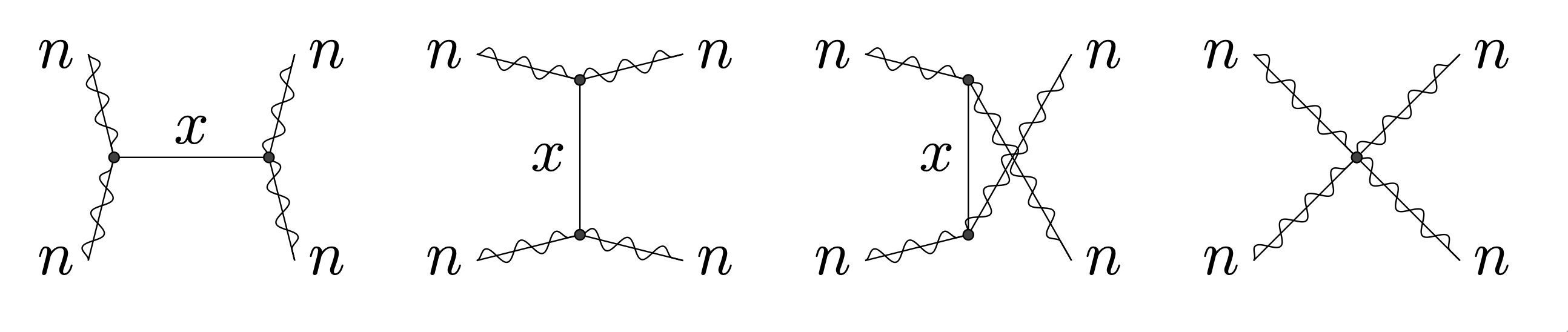}
    \caption{Feynman diagrams contributing to $n,n \to n,n$ massive spin-2 KK boson scattering. In a model incorporating the GW mechanism, the intermediate states $x$ include the radion, the graviton, massive spin-2 KK bosons, and GW scalars of various levels.}
    \label{fig:kk-scattering}
\end{figure}

The tree-level scattering amplitude receives contributions from a four-point contact interaction between the spin-2 modes, as well as from the exchange of intermediate spin-2 and spin-0 modes in the $s$-, $t$-, and $u$-channels as shown in Fig. \ref{fig:kk-scattering}. We write the total scattering amplitude as  
\begin{align}
    \mathcal{M} = \mathcal{M}_{c} +  \sum_{i=0}^{+\infty} \mathcal{M}_{(i)} + \sum_{j=0}^{+\infty} \mathcal{M}_{j}~,
    \label{eq:scattering-amplitude}
\end{align}
where $\mathcal{M}_{c}$ denotes the contribution from the contact interaction, $\mathcal{M}_{(i)}$ the contributions from the intermediate spin-0 states with KK number $i$, and $\mathcal{M}_{j}$ the contributions from the intermediate spin-2 states with KK number $j$.

\subsection{KK Mode Couplings}

In any extra-dimensional theory, an effective four-dimensional theory is attained by replacing each extra-dimensional field with a Kaluza-Klein decomposition in terms of four-dimensional KK modes and subsequently integrating over the extra dimensions. In this way, interactions between extra-dimensional fields yield couplings between four-dimensional states. For a single warped extra dimension, the procedure is explained in detail in \cite{Chivukula:2020hvi,Foren:2020egq} and, as explained there, the three- and four-point couplings between the spin-2 and spin-0 KK modes are attained from overlap integrals of products of corresponding extra-dimensional wavefunctions. In the stabilized case, these integrals are:
\begin{align}
    a_{lmn} &\equiv \dfrac{1}{\pi}\int_{-\pi}^{+\pi} d\varphi\hspace{10 pt}e^{-2A}\psi_{l}\,\psi_{m}\,\psi_{n}\\
    b_{l^{\prime}m^{\prime}n} &\equiv \dfrac{1}{\pi}\int_{-\pi}^{+\pi} d\varphi\hspace{10 pt}e^{-4A}(\partial_{\varphi}\psi_{l})(\partial_{\varphi}\psi_{m})\psi_{n}\\
    a_{klmn} &\equiv \dfrac{1}{\pi}\int_{-\pi}^{+\pi} d\varphi\hspace{10 pt}e^{-2A}\psi_{k}\,\psi_{l}\,\psi_{m}\,\psi_{n}\\
    b_{k^{\prime}l^{\prime}mn} &\equiv \dfrac{1}{\pi}\int_{-\pi}^{+\pi} d\varphi\hspace{10 pt}e^{-4A}(\partial_{\varphi}\psi_{k})(\partial_{\varphi}\psi_{l})\psi_{m}\,\psi_{n}\\
    a_{l^{\prime}m^{\prime}(n)} &\equiv \dfrac{1}{\pi}\int_{-\pi}^{+\pi} d\varphi\hspace{10 pt}e^{-2A}(\partial_{\varphi}\psi_{l})(\partial_{\varphi}\psi_{m})\gamma_{n}~.
\end{align}
Note that these couplings differ from one-another both in terms of warp factor dependence ($e^{-2A}$  versus $e^{-4A}$) and in which wavefunctions are differentiated.\footnote{In our previous work \cite{Chivukula:2019rij,Chivukula:2019zkt,Chivukula:2020hvi,Foren:2020egq} there was only a single massless radion mode; the coupling $b_{nnr}$ from that prior work has been relabeled as $a_{n^{\prime}n^{\prime}(0)}$ here.\label{fn:brnn}} As we have noted in prior work \cite{Chivukula:2020hvi,Foren:2020egq}, we can use the
spin-2 mode equation Eq. \eqref{eq:spin2-SLeqn} to rewrite
\begin{align}
    b_{l^{\prime}m^{\prime}n} = \dfrac{1}{2}\Big[\mu_{l}^{2} + \mu_{m}^{2} - \mu_{n}^{2}\Big] a_{lmn}\hspace{35 pt}b_{n^{\prime}n^{\prime}nn} = \dfrac{1}{3}a_{nnnn}~.
\end{align}
In what follows, we utilize these wherever possible.

\subsection{Scattering Amplitudes}

We now turn to the central results of this paper: reporting the scattering amplitude for elastic scattering of helicity-0, KK level-$n$, massive spin-2 KK bosons in a general GW model (as expressed in terms of the KK mode couplings), including the individual contributions from the four-point KK interaction, the exchange of spin-2 KK modes, and the exchange of spin-0 KK modes.
Our kinematic and helicity conventions follow those in \cite{Chivukula:2019rij,Chivukula:2019zkt,Chivukula:2020hvi,Foren:2020egq}. As in this prior work, we examine the scattering amplitudes for fixed center-of-mass scattering angle $\theta$ and examine the components which grow as different powers $s^\sigma$ of the center-of-mass 
energy-squared\footnote{In the helicity-0 amplitudes considered here, $\sigma$ is an integer only; however, other helicity combinations can yield half-integer powers of $s$ \cite{Chivukula:2020hvi,Foren:2020egq}.}
\begin{align}
    \mathcal{M}(s,\theta) = \sum_{\sigma \in \mathbb{Z}} \overline{\mathcal{M}}^{(\sigma)}(\theta)\, s^{\sigma}\hspace{35 pt}\text{ and }\hspace{35 pt}\mathcal{M}^{(\sigma)}\equiv \overline{\mathcal{M}}^{(\sigma)}s^{\sigma}~.
    \label{eq:powerexpansion}
\end{align}
In the subsections which follow, we compute the contributions to the scattering amplitude at each order ($\overline{\mathcal{M}}^{(\sigma)}(\theta)$ for $\sigma\in\{2,3,4,5\}$) and isolate the sum rule relationships between the couplings and masses that must be satisfied so that the contributions cancel at each order. We summarize these sum rules in the last subsection.

\subsubsection{$\mathcal{O}(s^{5})$ and $\mathcal{O}(s^{4})$}

The only contributions to the scattering amplitude at $\mathcal{O}(s^{5})$ and $\mathcal{O}(s^{4})$ come from the contact interaction and exchange of intermediate spin-2 modes \cite{ArkaniHamed:2002sp,ArkaniHamed:2003vb,Chivukula:2019zkt,Bonifacio:2019ioc}. Therefore, the form of the scattering amplitude at these orders and the resulting sum rules are unchanged from \cite{Chivukula:2019rij,Bonifacio:2019ioc}. We include them here for completeness.

At order $s^5$, we find the following contributions:
\begin{align}
    \mathcal{M}_{j}^{(5)} &=  \dfrac{\kappa^{2}s^{5}r_{c}^{7}}{2304 \pi \mu_{n}^{8}} \bigg\{a_{nnj}^{2}\bigg\}\,(7+c_{2\theta})\,s_{\theta}^{2}~,\\
     \mathcal{M}_{(i)}^{(5)} &= 0~,\\
    \mathcal{M}_{c}^{(5)} &= \dfrac{\kappa^{2}s^{5}r_{c}^{7}}{2304 \pi \mu_{n}^{8}} \bigg\{-a_{nnnn}\bigg\}\,(7+c_{2\theta})\,s_{\theta}^{2}~.
\end{align}
Combining these, we get
\begin{align}
    \mathcal{M}^{(5)} &= \mathcal{M}_{c}^{(5)} + \sum_{j=0}^{+\infty} \mathcal{M}_{j}^{(5)} + \sum_{i=0}^{+\infty} \mathcal{M}_{(i)}^{(5)}\nonumber\\
    &= \dfrac{\kappa^{2}s^{5}r_{c}^{7}}{2304 \pi \mu_{n}^{8}} \bigg\{ -a_{nnnn} + \sum_{j= 0} a_{nnj}^{2} \bigg\}\,(7+c_{2\theta})\,s_{\theta}^{2}~.
    \label{eq:orders5}
\end{align}
The ${\mathcal O}(s^5)$ contributions collectively cancel only if the coupling quantity in curly brackets vanishes.  That vanishing defines the  ${\mathcal O}(s^5)$ sum rule \cite{Chivukula:2019zkt,Chivukula:2020hvi,Foren:2020egq}.

Next, at order $s^4$, we find
\begin{align}
    \mathcal{M}_{j}^{(4)} &=  \dfrac{\kappa^{2}s^{4}r_{c}^{5}}{27648 \pi \mu_{n}^{8}} \bigg\{-3\,a_{nnj}^{2}\bigg[(7+c_{2\theta})^{2}\mu_{j}^{2} + 2(9-140c_{2\theta}+3c_{4\theta})\mu_{n}^{2}\bigg]\bigg\}~,\\
     \mathcal{M}_{(i)}^{(4)} &= 0~,\\
    \mathcal{M}_{c}^{(4)} &= \dfrac{\kappa^{2}s^{4}r_{c}^{5}}{27648 \pi \mu_{n}^{8}} \bigg\{4\,a_{nnnn}\bigg[ 63 - 196c_{2\theta} + 5c_{4\theta} \bigg]\mu_{n}^{2}\bigg\}~.
\end{align}
Applying the $\mathcal{O}(s^{5})$ sum rule to simplify their sum, we find
\begin{align}
    \mathcal{M}^{(4)} &= \mathcal{M}_{c}^{(4)} + \sum_{j=0}^{+\infty} \mathcal{M}_{j}^{(4)} + \sum_{i=0}^{+\infty} \mathcal{M}_{(i)}^{(4)}\nonumber\\
    &= \dfrac{\kappa^{2}s^{4}r_{c}^{5}}{27648 \pi \mu_{n}^{8}} \bigg\{ 4 \mu_{n}^{2} a_{nnnn} - 3 \sum_{j= 0}^{+\infty} \mu_{j}^{2} a_{nnj}^{2} \bigg\}\,(7+c_{2\theta})~.
       \label{eq:orders4}
\end{align}
The vanishing of the coupling combination found in the curly brackets above defines the ${\mathcal O}(s^4)$ sum rule \cite{Chivukula:2019zkt,Chivukula:2020hvi,Foren:2020egq}.

Because the scalar tower is not relevant at $\mathcal{O}(s^{5})$ and $\mathcal{O}(s^{4})$, these sum rules are identical to those quoted in \cite{Chivukula:2019zkt,Chivukula:2020hvi,Foren:2020egq}.
As demonstrated in \cite{Chivukula:2019zkt,Bonifacio:2019ioc}, the orthogonality and completeness of the spin-2 modes -- which is guaranteed by the Sturm-Liouville form of the spin-2 mode equation -- is sufficient to prove these sum rules are satisfied in any model with the structure we described. In particular, the sum rules are satisfied for any background warp function $A(y)$ and are satisfied in any model incorporating the GW mechanism which produces the required stabilized geometry.

\subsubsection{$\mathcal{O}(s^{3})$}
At order $s^3$, we find the following contributions:
\begin{align}
    \mathcal{M}_{0}^{(3)} &=  \dfrac{\kappa^{2}s^{3}r_{c}^{3}}{6912 \pi \mu_{n}^{8}} \bigg\{6\,a_{nn0}^{2}\bigg[15 - 270 c_{2\theta} - c_{4\theta}\bigg]\mu_{n}^{4}\bigg\}~,\\
    \mathcal{M}_{j}^{(3)} &=  \dfrac{\kappa^{2}s^{3}r_{c}^{3}}{6912 \pi \mu_{n}^{8}} \bigg\{3\,a_{nnj}^{2}\bigg[(10 s_{\theta}^{2})\mu_{j}^{4} + (69+60 c_{2\theta} - c_{4\theta}) \mu_{n}^{2}\mu_{j}^{2}\nonumber\\
    &\hspace{35 pt}+ 2(13-268c_{2\theta}-c_{4\theta})\mu_{n}^{4}\bigg]\bigg\}~,\\
     \mathcal{M}_{(i)}^{(3)} &= \dfrac{\kappa^{2}s^{3}r_{c}^{3}}{6912 \pi \mu_{n}^{8}}\bigg\{-216\, a_{n^{\prime}n^{\prime}(i)}^{2} s_{\theta}^{2}\bigg\}~,\\
    \mathcal{M}_{c}^{(3)} &= \dfrac{\kappa^{2}s^{3}r_{c}^{3}}{6912 \pi \mu_{n}^{8}} \bigg\{2 \,a_{nnnn}\bigg[ -185 + 692c_{2\theta} + 5c_{4\theta} \bigg]\mu_{n}^{4}\bigg\}~.
\end{align}
Summing these, and applying the $\mathcal{O}(s^{5})$ and $\mathcal{O}(s^{4})$ sum rules, we find
\begin{align}
    \mathcal{M}^{(3)} &= \mathcal{M}_{c}^{(3)} + \sum_{j=0}^{+\infty} \mathcal{M}_{j}^{(3)} + \sum_{i=0}^{+\infty} \mathcal{M}_{(i)}^{(3)}\nonumber\\
    &\hspace{-10 pt}= \dfrac{\kappa^{2}s^{3}r_{c}^{3}}{3456 \pi \mu_{n}^{8}} \bigg\{ 4\mu_{n}^{4}(3\,a_{nn0}^{2} - 4\,a_{nnnn})-108 \sum_{i=0}^{\infty} a_{n^{\prime}n^{\prime}(i)}^{2} +15 \sum_{j=0}^{+\infty} \mu_{j}^{4} a_{nnj}^{2} \bigg\}\,s_{\theta}^{2}~.
       \label{eq:orders3}
\end{align}
At this order, the scalars (through the couplings $a_{n'n'(i)}$) begin to yield nonzero contributions \cite{ArkaniHamed:2002sp,ArkaniHamed:2003vb,Chivukula:2019zkt,Bonifacio:2019ioc}. The vanishing of the combination of couplings in the curly brackets above generalizes  the ${\mathcal O}(s^3)$ sum rule found previously \cite{Chivukula:2019zkt,Chivukula:2020hvi,Foren:2020egq}. In particular, this new $\mathcal{O}(s^{3})$ sum rule contains contributions from the exchange of GW scalars: the results at ${\mathcal O}(s^3)$ in the unstabilized RS model \cite{Chivukula:2019zkt} amount to truncating the scalar tower sum to a single massless radion (see footnote 10).

\subsubsection{$\mathcal{O}(s^{2})$}
Finally, at order $s^2$ we find
\begin{align}
    \mathcal{M}_{0}^{(2)} &=  \dfrac{\kappa^{2}s^{2}r_{c}}{6912 \pi \mu_{n}^{8}} \bigg\{12\,a_{nn0}^{2}\bigg[175+624 c_{2\theta} + c_{4\theta}\bigg]\mu_{n}^{6}\bigg\}~,\\
    \mathcal{M}_{j}^{(2)} &=  \dfrac{\kappa^{2}s^{2}r_{c}}{6912 \pi \mu_{n}^{8}} \bigg\{a_{nnj}^{2}\bigg[-8(7+c_{2\theta})\mu_{j}^{6} + 20(7+c_{2\theta})\mu_{n}^{2}\mu_{j}^{4}\nonumber\\
    &\hspace{35 pt}- (1291+1132 c_{2\theta} +9 c_{4\theta}) \mu_{n}^{4}\mu_{j}^{2} + 4(553+1876c_{2\theta}+3c_{4\theta})\mu_{n}^{6}\bigg]\bigg\}~,\\
     \mathcal{M}_{(i)}^{(2)} &= \dfrac{\kappa^{2}s^{2}r_{c}}{6912 \pi \mu_{n}^{8}}\bigg\{72\, a_{n^{\prime}n^{\prime}(i)}^{2} (7+c_{2\theta})\bigg[2 \mu_{n}^{2} - \mu_{(i)}^{2}\bigg]\bigg\}~,\\
    \mathcal{M}_{c}^{(2)} &= \dfrac{\kappa^{2}s^{2}r_{c}}{6912 \pi \mu_{n}^{8}} \bigg\{-128\, a_{nnnn}\bigg[ 5 + 47 c_{2\theta} \bigg]\mu_{n}^{6}\bigg\}~.
\end{align}
Applying the $\mathcal{O}(s^{5})$, $\mathcal{O}(s^{4})$, and newly generalized $\mathcal{O}(s^{3})$ sum rules, we find
\begin{align}
    \mathcal{M}^{(2)} &= \mathcal{M}_{c}^{(2)} + \sum_{j=0}^{+\infty} \mathcal{M}_{j}^{(2)} + \sum_{i=0}^{+\infty} \mathcal{M}_{(i)}^{(2)}\nonumber\\
    &\hspace{-10 pt}= -\dfrac{\kappa^{2}s^{2}r_{c}}{864 \pi \mu_{n}^{8}} \bigg\{ 4 \mu_{n}^{6} a_{nn0}^{2} +9 \sum_{i=0}^{+\infty} (\mu_{(i)}^{2} - 4 \mu_{n}^{2}) a_{n^{\prime}n^{\prime}(i)}^{2}+ \sum_{j=0}^{+\infty} \mu_{j}^{6} a_{nnj}^{2} \bigg\}\,(7+c_{2\theta})~.
       \label{eq:orders2}
\end{align}
The vanishing of the expression in the curly brackets above generalizes the ${\mathcal O}(s^2)$ sum rule of \cite{Chivukula:2019zkt,Chivukula:2020hvi,Foren:2020egq}. At this order, the masses of all scalar modes contribute directly. In particular, not only do the massive GW scalars contribute, but so does the mass of the radion, $\mu^2_{(0)}$, which vanished in the unstabilized model.

\subsubsection{$\mathcal{O}(s)$}

Following the cancellations at $\mathcal{O}(s^{5})$ through $\mathcal{O}(s^{2})$, the overall $n,n\rightarrow n,n$ scattering amplitude of helicity-0, KK level-$n$, massive spin-2 KK bosons grows only like ${\mathcal O}(s)$ at high energies. After applying the sum rules we isolated above, we find the leading high energy behavior of the amplitude is

\begin{align}
    \mathcal{M}^{(1)} &= \dfrac{\kappa^{2}s}{34560\pi r_{c}\, \mu_{n}^{8}} \bigg\{4\mu_{n}^{8}(36\, a_{nn0}^{2} + 7\,a_{nnnn}) - 216 \sum_{i=0}^{+\infty} \mu_{n}^{2}(6\mu_{n}^{2}-5\mu_{(i)}^{2})a_{n^{\prime}n^{\prime}(i)}^{2}\nonumber\\
    &\hspace{35 pt} + 15 \sum_{j=0}^{+\infty} \mu_{j}^{8} a_{nnj}^{2} \bigg\} \dfrac{(7+c_{2\theta})^{2}}{ s_{\theta}^{2}}~.
\end{align}
This generalizes the result that was given previously in \cite{Chivukula:2020hvi,Foren:2020egq}, by now including contributions from the entire scalar tower.

\subsection{Sum Rules}

\label{subsec:sum-rules}

To summarize, by requiring the scattering amplitude to grow no faster than $\mathcal{O}(s)$ in the GW model, we determine the following general sum rules should be satisfied:
\begin{align}
    \sum_{j=0} a_{nnj}^{2} &= a_{nnnn}~, \label{eq:sr1}\\
    \sum_{j=0} \mu_{j}^{2} a_{nnj}^{2} &= \dfrac{4}{3} \mu_{n}^{2} a_{nnnn}~,\label{eq:sr2}\\
    \sum_{j=0}^{+\infty} \mu_{j}^{4} a^{2}_{nnj} &=  \dfrac{4}{15}\mu_{n}^{4} (4\,a_{nnnn} - 3\,a^{2}_{nn0}) + \dfrac{36}{5} \sum_{i=0}^{+\infty} a^{2}_{n^{\prime}n^{\prime}(i)}~,\label{eq:sr3}\\
    \sum_{j=0}^{+\infty} \mu_{j}^{6} a^{2}_{nnj} &= -4\mu_{n}^{6}a^{2}_{nn0} + 9 \sum_{i=0}^{+\infty}(4\mu_{n}^{2}-\mu_{(i)}^{2}) a^{2}_{n^{\prime}n^{\prime}(i)}~, \label{eq:sr4}
\end{align}
The first two sum rules, Eqs. \eqref{eq:sr1} and \eqref{eq:sr2}, follow from the Sturm-Liouville form of the spin-2 KK mode equation, Eq. \eqref{eq:spin0mode-equation} - so the proofs given in \cite{Chivukula:2019zkt,Chivukula:2020hvi,Foren:2020egq} apply to any model producing a geometry defined by a warp function $A(y)$. The sum rules in Eqs. \eqref{eq:sr3} and \eqref{eq:sr4}, however, are new - they involve the scalar tower newly present in the GW model. 

By combining the last two sum rules, Eqs. \eqref{eq:sr3} and \eqref{eq:sr4}, to eliminate the common sum $\sum_{i} a_{n^{\prime}n^{\prime}(i)}^{2}$, we find a mixed rule:
\begin{align}
    \sum_{j=0}^{+\infty} \bigg[5\mu_{n}^{2}-\mu_{j}^{2}\bigg]\mu_{j}^{4} a^{2}_{nnj} &= \dfrac{16}{3}\mu_{n}^{6}a_{nnnn} + 9\sum_{i=0}^{+\infty}\mu_{(i)}^{2} a_{n^{\prime}n^{\prime}(i)}^{2} ~.\label{eq:sr5}
\end{align}
The only scalar tower sum $\sum_{i=0}^{+\infty} \mu_{(i)}^{2} a^{2}_{n^{\prime}n^{\prime}(i)}$ remaining in this particular combination of the $\mathcal{O}(s^{3})$ and $\mathcal{O}(s^{2})$ sum rules can be eliminated via properties of the spin-0 wavefunctions. In particular, using the spin-0 mode equation, Eq. \eqref{eq:spin0mode-equation}; the corresponding normalizations, Eqs. \eqref{eq:scalarnorm1} and \eqref{eq:scalarnorm2}; and assuming completeness, we can prove that the mixed sum rule Eq. \eqref{eq:sr5} holds in any GW model. Details and further discussion will be given in \cite{Chivukula:2020tbd}. 

In the next section, we introduce the flat stabilized model, which implements the GW mechanism and in which we can directly verify the sum rules.
\section{The Flat Stabilized Model}

\label{sec:flat-stabilized-model}

DeWolfe, Freedman, Gubser, and Karch (DFGK) \cite{DeWolfe:1999cp} have given a general prescription for producing solutions to the coupled bulk scalar and five-dimensional gravity field equations within theories that implement the Goldberger-Wise \cite{Goldberger:1999uk, Goldberger:1999un} mechanism. Using this prescription, we can build self-consistent models having a stabilized extra dimension. In this section, we begin by reviewing the DFGK construction. We then define the flat stabilized model and the (small) parameter which deforms the geometry away from a five-dimensional orbifolded toroidal space. We subsequently compute the wavefunctions and masses of the spin-0 and spin-2 KK modes. Finally, we compute the KK mode couplings and demonstrate that the sum rules defined above are satisfied to leading nontrivial order in the deformation parameter.

\subsection{The DFGK Model}

The DFGK model is determined in terms of a superpotential $W[\hat{\phi}]$. The authors of \cite{DeWolfe:1999cp} demonstrate that $W$ guarantees self-consistent background solutions to the coupled gravity-scalar theory if
\begin{align}
       (\partial_{\varphi} A) = \frac{W}{12}\bigg|_{\hat{\phi}=\phi_{0}}\text{ sign}(\varphi)\hspace{20 pt}&\hspace{20 pt}(\partial_{\varphi} \phi_{0}) = \dfrac{dW}{d\hat{\phi}}\bigg|_{\hat{\phi}=\phi_{0}}\text{ sign}(\varphi)~,\label{AAndPhiInTermsOfW}
\end{align}
and, for the bulk and brane potentials,
\begin{align}
        &\hspace{-60 pt}V r_{c}^{2} = \dfrac{1}{8}\bigg(\dfrac{dW}{d\hat{\phi}}\bigg)^{2} - \dfrac{W^2}{24}\\
    V_{1}r_{c} = +\dfrac{W}{2} + \beta_{1}^{2}\bigg[\hat{\phi}(\varphi) -\phi_{1}\bigg]^2\hspace{20 pt}&\hspace{20 pt}V_{2}r_{c} = -\dfrac{W}{2} + \beta_{2}^{2}\bigg[\hat{\phi}(\varphi)-\phi_{2}\bigg]^2~. \label{eq:branepotentials}
\end{align}
 where $\phi_{1} \equiv \hat{\phi}(0)$ and $\phi_{2} \equiv \hat{\phi}(\pi)$. In particular, for the choice\footnote{Here $u$ is a parameter, and not the $\hat{u}$ field of Eq. \eqref{eq:uhat}.} \cite{DeWolfe:1999cp} 
\begin{align}
    W[\hat{\phi}(\varphi)] = 12 kr_{c} - \dfrac{1}{2}\hat{\phi}(\varphi)^{2}\,ur_c~,
\end{align}
Eq. \eqref{AAndPhiInTermsOfW} become exactly solvable, satisfied by background solutions
\begin{align}
    \phi_0(\varphi) &= \phi_{1}e^{-ur_c|\varphi|} \label{eq:scalarbackgroundi}\\
    A(\varphi) &= kr_{c}|\varphi| + \dfrac{1}{48}\phi_{1}^{2}\bigg[e^{-2ur_c|\varphi|} - 1\bigg]
\end{align}
where, consistent with Eq. \eqref{eq:branepotentials},
\begin{equation}
    ur_c=\dfrac{1}{\pi}\log\dfrac{\phi_1}{\phi_2}~.
\end{equation}

For small $ur_c$, we find
\begin{align}
    A(\varphi) &= kr_{c} |\varphi| - \bigg[\dfrac{\phi_{1}^{2}}{24}|\varphi|\bigg](ur_c) + \bigg[\dfrac{\phi_{1}^{2}}{24}|\varphi|^{2}\bigg](ur_c)^{2} + \mathcal{O}\Big((ur_c)^{3}\Big)\\
    &= \bigg[k - \dfrac{\phi_{1}^{2}u}{24}  \bigg]r_{c}|\varphi| + \bigg[\dfrac{\phi_{1}^{2}(ur_c)^{2}}{24} \bigg] |\varphi|^{2} + \mathcal{O}\Big((ur_c)^{3}\Big)~,
\end{align}
and thus we have a stabilized model which is a small (in $ur_c$) deformation of the usual Randall-Sundrum model \cite{Randall:1999ee,Randall:1999vf} with an effective warp parameter $\tilde{k} \equiv k - \phi_{1}^{2}u/24 $ \cite{DeWolfe:1999cp}.

\subsection{The Flat Stabilized Limit}

Inspired by this, we rewrite the DFGK model in terms of $\tilde{k}$ and replace the dimensionless small parameter $ur_c$ with a new dimensionless small parameter $\epsilon\equiv \phi_{1}(ur_c)/\sqrt{24}$. This yields, to all orders in $\epsilon$,
\begin{align}
    W[\hat{\phi}(\varphi)] &= 12 \tilde{k}r_{c} + \dfrac{\sqrt{6}}{\phi_{1}} \bigg[\phi_{1}^{2} -\hat{\phi}(\varphi)^{2}\bigg]\epsilon~,\\
    \phi_{0}(\varphi) &= \phi_{1} \exp\bigg(-\dfrac{2\sqrt{6}}{\phi_{1}}\epsilon|\varphi|\bigg)~, \label{eq:scalarbackground}
\end{align}
and
\begin{align}
    A(\varphi) &= \tilde{k}r_{c}|\varphi|+ \dfrac{\phi_{1}^{2}}{48}\bigg[\exp\bigg(-\dfrac{4\sqrt{6}}{\phi_{1}}\epsilon|\varphi|\bigg)-1\bigg] + \dfrac{\phi_{1}}{2\sqrt{6}}\epsilon|\varphi|~,\\
    &= \tilde{k}r_{c}|\varphi| + \epsilon^{2}|\varphi|^{2} + \mathcal{O}(\epsilon^{3})~.
\end{align}

The flat stabilized model is the $\tilde{k}\to 0$ limit of the above theory, in which the warp factor becomes
\begin{align}
    A(\varphi) &= \epsilon^{2}|\varphi|^{2} + \mathcal{O}(\epsilon^{3})
    \label{eq:flat-stabilized-A}
\end{align}
and the scalar background is given by Eq. \eqref{eq:scalarbackground}. 
When $\epsilon=0$, this theory reduces to the (unstabilized) five-dimensional orbifold torus model, a model we have previously analyzed with respect to massive spin-2 KK mode scattering \cite{Chivukula:2019rij}. The flat stabilized model is therefore a deformation (in $\epsilon$) of a flat unstabilized model.

\subsection{Wavefunctions and Eigenvalues}

\renewcommand{\arraystretch}{2}
\begin{table}
\begin{center}
 \begin{tabular}{ | c | c || c | c | c |}
    \hline
    & & $\mathcal{O}(1)$ & $\mathcal{O}(\epsilon)$ & $\mathcal{O}(\epsilon^{2})$\\[0.5 em]
    \hline\hline
    graviton wfxn & $\psi_{0}$ & $\frac{1}{\sqrt{2}}$ & - & $\frac{\pi^{2}}{3\sqrt{2}}$\\[0.5 em]
    \hline
    graviton mass$^{2}$ & $\mu_{0}^{2}$ & - & - & -\\[0.5 em]
    \hline\hline
    spin-2 KK wfxn & $\psi_{n}$ & $-c_{n\varphi}$ & -& \centered{$\frac{1}{6n^{2}}\Big[9 + n^{2}(\pi^{2}-9\varphi^{2})\Big]c_{n\varphi}$\\ $\hspace{10 pt}+ \frac{\varphi}{3n}\Big[9 + n^{2}(\varphi^{2} -\pi^{2})\Big]s_{n\varphi}$\\[0.5 em]} \\[0.5 em]
    \hline
    spin-2 KK mass$^{2}$ & $\mu_{n}^{2}$ & $n^{2}$ & - & $3-\frac{2n^{2}\pi^{2}}{3}$\\[0.5 em]
    \hline\hline
    radion wfxn & $\gamma_{0}$ & $\frac{1}{\sqrt{2}}$ & - & $-\frac{\pi^{2}}{3\sqrt{2}}$\\[0.5 em]
    \hline
    radion mass$^{2}$ & $\mu_{(0)}^{2}$ & - & - & $4$\\[0.5 em]
    \hline\hline
    GW scalar wfxn & $\gamma_{n}$ & - & $-\frac{2}{n} c_{n\varphi}$ & $\frac{4\sqrt{6}}{n^{2}\phi_{1}}\Big[(n\varphi) c_{n\varphi} - s_{n\varphi}\Big]$\\[0.5 em]
    \hline
    GW scalar mass$^{2}$ & $\mu_{(n)}^{2}$ & $n^{2}$ & - & $1 - \frac{2n^{2}\pi^{2}}{3} + \frac{24}{\phi_{1}^{2}}$\\[0.5 em]
    \hline
\end{tabular}
\caption{{\bf Wavefunctions and Eigenvalues:} Perturbative results for the spin-2 and spin-0 KK wavefunctions and masses up to ${\mathcal O}(\epsilon^2)$ in the flat stabilized model. The leading-order results for the spin-2 KK masses correspond to those of the five-dimensional toroidal orbifold model used in \cite{Chivukula:2019rij}. The radion mass $\mu^{2}_{(0)}$ starts at order ${\mathcal O}(\epsilon^2)$ \cite{Tanaka:2000er,Csaki:2000zn}. Due to the normalization conditions of Eqs. \eqref{eq:scalarnorm1} and \eqref{eq:scalarnorm2}, the radion wavefunction $\gamma_0$ starts at ${\mathcal O}(1)$, whereas the GW scalar wavefunctions $\gamma_{n}$ start at ${\mathcal O}(\epsilon)$. Here $c_{n\varphi}\equiv \cos(n\varphi)$ and $s_{n\varphi}\equiv\sin(n\varphi)$, and ``wfxn" stands for wavefunction. At ${\mathcal O}(1)$ these results agree with those found in \cite{Chivukula:2019rij}.}
\label{table:wavefunctions}
\end{center}
\end{table}
\renewcommand{\arraystretch}{1}

Next, we perturbatively compute all wavefunctions and masses-squared of the spin-2 and spin-0 KK modes in the flat stabilized model to $\mathcal{O}(\epsilon^{2})$. Before quoting our results, let us first understand the form of the results we should expect. In the spin-2 sector, the analysis is relatively straightforward: the mode equation and normalization conditions, Eqs. \eqref{eq:spin2-SLeqn} and \eqref{eq:spin2-mode-norm} behave smoothly as $\epsilon \to 0$, and the usual expectations from Rayleigh-Schr\"odinger perturbation theory apply.

The spin-0 sector is trickier. First we use the stiff-wall limit defined in Eq. \eqref{eq:stiff-wall} to simplify the scalar boundary conditions and to ensure that the scalar system defines a Sturm-Liouville problem. Next, we note that Eq. \eqref{eq:scalarbackground} directly implies that the quantity $(\partial_\varphi \phi_0)^2$ present in the denominators of either form of the scalar normalization condition, Eqs. \eqref{eq:scalarnorm1} and \eqref{eq:scalarnorm2}, is an $\mathcal{O}(\epsilon^{2})$ quantity, thereby yielding divergences in the $\epsilon\rightarrow 0$ limit if not handled with care.\footnote{The limit $\epsilon\to 0$ corresponds to the ``unstabilized" limit discussed at the end of Sec. \ref{sec:spin-0-modes}.}

In the case of the radion, we know on general grounds \cite{Tanaka:2000er,Csaki:2000zn} that the mass-squared of the radion must vanish if one neglects the backreaction of the scalar field configuration on the geometry. As we demonstrate explicitly,  the mass-squared of the radion begins at ${\mathcal O}(\epsilon^2)$: computing in perturbation theory, we find that
\begin{equation}
    \mu^2_{(0)}=4 \epsilon^2 + \mathcal{O}(\epsilon^{3})~.
\end{equation} 
Since the radion mass $\mu^2_{(0)}$ is ${\mathcal O}(\epsilon^2)$ to leading order in $\epsilon$,  the radion wavefunction $\gamma_{0}$ must be ${\mathcal O}(1)$ in order to be consistent with the spin-0 normalization condition as written in Eq. \eqref{eq:scalarnorm2}.

By contrast, the GW scalar masses $\mu^2_{(i)}$ are ${\mathcal O}(1)$ to leading order in $\epsilon$, and therefore the normalization conditions, Eqs. \eqref{eq:scalarnorm1} and \eqref{eq:scalarnorm2}, imply that the GW scalar wavefunctions $\gamma_{i}$ must be ${\mathcal O}(\epsilon)$. Physically, the fact that each GW scalar wavefunction $\gamma_{(i)}$ vanishes as $\epsilon \to 0$ reflects the fact that the GW scalars decouple from the scalar metric sector in the unstabilized limit.

In order to verify the sum rules Eqs. \eqref{eq:sr1} - \eqref{eq:sr4} to nontrivial order, we need wavefunctions and masses of the spin-2 and spin-0 modes to $\mathcal{O}(\epsilon^{2})$. These can be calculated by applying the defining equations of the flat stabilized model, Eqs. \eqref{eq:flat-stabilized-A} and \eqref{eq:scalarbackground}, to the appropriate mode equations and normalization conditions. For the spin-2 modes, these are Eqs. \eqref{eq:spin2-SLeqn} and \eqref{eq:spin2-mode-norm} respectively. For the spin-0 modes, these are either Eqs. \eqref{eq:spin0mode-equation} and \eqref{eq:scalarnorm1} or Eqs. \eqref{eq:spin0mode-equation} and \eqref{eq:scalarnorm2}, where the former (latter) normalization condition is more useful for the radion mode (GW scalar modes). The results of this perturbative calculation are summarized in Table \ref{table:wavefunctions}, while details of the calculation are
supplied in Appendix \ref{sec:appendix}.\footnote{Because the sum rules only involve the GW scalar mass in the combination $\mu_{(n)}^{2} \gamma_{n}^{2}$ and the GW scalar wavefunctions $\gamma_{n}$ are $\mathcal{O}(\epsilon)$, we actually only need the GW scalar mass-squared to $\mathcal{O}(1)$. Its $\mathcal{O}(\epsilon^{2})$ contribution is included in Table \ref{table:wavefunctions} for completeness.}

We next compute the couplings between these modes relevant to the sum rules.

\subsection{KK Mode Couplings and Sum Rules}

In the flat stabilized model, we can directly evaluate the integrals present in the couplings relevant to the sum rules through $\mathcal{O}(\epsilon^{2})$. Using the wavefunctions in Table \ref{table:wavefunctions}, we find
\begin{align}
    a_{nnj} = \begin{cases}
    j=0:\hspace{10 pt}\dfrac{1}{\sqrt{2}} + \bigg[\dfrac{\pi^{2}}{3\sqrt{2}}\bigg]\epsilon^{2}\\
    j=2n:\hspace{10 pt}-\dfrac{1}{2} + \bigg[-\dfrac{27}{32n^{2}} -\dfrac{\pi^{2}}{6}\bigg]\epsilon^{2}\\
    \text{else}:\hspace{10 pt}\bigg[-\dfrac{96(-1)^{j}n^{4}}{(j^{3}-4jn^{2})^{2}}\bigg]\epsilon^{2}\\
    \end{cases}
\end{align}

\begin{align}
    a_{n^{\prime}n^{\prime}(i)} = \begin{cases}
    i=0:\hspace{10 pt}\dfrac{n^{2}}{\sqrt{2}} + \bigg[\dfrac{6-\pi^{2}n^{2}}{3\sqrt{2}}\bigg]\epsilon^{2}\\
    i=2n:\hspace{10 pt}\bigg[\dfrac{n}{2}\bigg]\epsilon + \bigg[-\sqrt{\dfrac{3}{2}} \dfrac{\pi n}{\phi_{1}}\bigg]\epsilon^{2}\\
    \text{else}:\hspace{10 pt}\bigg[\dfrac{64\sqrt{6}[(-1)+(-1)^{i}]n^{4}[2n^{2}-i^{2}]}{i^{3}(i^{2}-4n^{2})^{2}}\bigg]\epsilon^{2}\\
    \end{cases}
\end{align}

\begin{align}
    a_{nnnn} = \dfrac{3}{4} + \bigg[\dfrac{27}{32n^{2}} + \dfrac{\pi^{2}}{2}\bigg]\epsilon^{2}
\end{align}

At zeroth order in $\epsilon$, the theory inherits the discrete momentum conservation present in the five-dimensional orbifold torus model, and only the level-0 modes (the graviton and radion) and the level-$2n$ modes (specific massive spin-2 and GW scalar states) contribute. At higher orders in $\epsilon$, the background scalar field breaks discrete momentum conservation; however, as the above computations demonstrate, any contributions to the scattering amplitude from the exchange of modes at levels other than $0$ and $2n$ are suppressed beyond ${\mathcal O}(\epsilon^2)$. Therefore, for the purposes of calculating the scattering amplitude to $\mathcal{O}(\epsilon^{2})$, we need only include corrections to masses and couplings relating to the radion, graviton, level-$2n$ massive spin-$2$ KK mode, and the level-$2n$ GW scalar.

\renewcommand{\arraystretch}{2}
\begin{table}
\begin{center}
\begin{tabular}{|| c || c | c ||}
\hline
& $\mathcal{O}(1)$ & $\mathcal{O}(\epsilon^{2})$\\[0.5 em]
\hline\hline
$a_{nn0}^{2}$ & $\frac{1}{2}$ & $\frac{\pi^{2}}{3}$ \\[0.5 em]
\hline\hline
$\sum_{j\neq 0} a_{nnj}^{2}$ & $\frac{1}{4}$ & $\frac{27}{32n^{2}} + \frac{\pi^{2}}{6}$\\[0.5 em]
\hline
$\sum_{j=0} \mu_{j}^{2p}\,a_{nnj}^{2}$ & $\frac{n^{2p}}{4^{1-p}}$ & $\frac{2^{2p-5}}{3n^{2-2p}} \Big[9(9+2p) + 16(1-p)n^{2}\pi^{2}\Big]$\\[0.5 em]
\hline\hline
$a_{n^{\prime}n^{\prime}(0)}^{2}$ & $\frac{n^{4}}{2}$ & $2n^{2} - \frac{n^{4}\pi^{2}}{3}$ \\[0.5 em]
\hline
$\mu_{(0)}^{2} a_{n^{\prime}n^{\prime}(0)}^{2}$ & - & $2n^{4}$ \\[0.5 em]
\hline\hline
$\sum_{i\neq 0} a_{n^{\prime}n^{\prime}(i)}^{2}$ & - & $\frac{n^{2}}{4}$\\[0.5 em]
\hline
$\sum_{i\neq 0} \mu_{(i)}^{2p}a_{n^{\prime}n^{\prime}(i)}^{2}$ & - & $\frac{n^{2+2p}}{4^{1-p}}$\\[0.5 em]
\hline\hline
$a_{nnnn}$ & $\frac{3}{4}$ & $\frac{27}{32n^{2}}+\frac{\pi^{2}}{2}$\\[0.5 em]
\hline
$\mu_{n}^{2p}a_{nnnn}$ & $\frac{3n^{2p}}{4}$ & $\frac{1}{32n^{2-2p}}\Big[9(3+8p)+16(1-p)n^{2}\pi^{2}\Big]$\\[0.5 em]
\hline
\end{tabular}
\caption{{\bf Sum Rule Coupling Combinations:} This table lists the values of the combinations of couplings that appear in the sum rules -- Eqs. \eqref{eq:sr1}, \eqref{eq:sr2}, \eqref{eq:sr3}, and \eqref{eq:sr4} -- which ensure that the  $n,n\rightarrow n,n$ scattering amplitude of helicity-0, KK level-$n$, massive spin-2 KK bosons grows no faster than $s$  in the flat stabilized model including contributions through ${\mathcal O}(\epsilon^2)$.  At ${\mathcal O}(1)$, these results agree with those found in \cite{Chivukula:2019rij}.}
\label{table:coupling-combinations}
\end{center}
\end{table}
\renewcommand{\arraystretch}{1}

In Table \ref{table:coupling-combinations}, we report the combinations of these couplings which arise in our computation of the $n,n \to n,n$ scattering amplitude. 
Using these results we verify all of the sum rules through ${\mathcal O}(\epsilon^2)$. As we discuss below, all contributions identified above are essential at ${\mathcal O}(\epsilon^2)$ in demonstrating the cancellations necessary for an $\mathcal{O}(s)$ scattering amplitude.

Neither radion nor GW scalar exchange contributes directly to the scattering amplitude at ${\mathcal O}(s^5)$ and ${\mathcal O}(s^4)$ \cite{ArkaniHamed:2002sp,ArkaniHamed:2003vb}. However, the stabilization procedure does impact the sum rules in a less obvious way: for the $\mathcal{O}(s^{5})$ and $\mathcal{O}(s^{4})$ sum rules, Eqs. \eqref{eq:sr1} and \eqref{eq:sr2}, there are ${\mathcal O}(\epsilon^2)$ contributions to the spin-2 KK mode wavefunctions, masses, and couplings coming from the deformed geometry. As noted previously, the cancellations of $\mathcal{O}(s^{5})$ and $\mathcal{O}(s^{4})$ contributions to the scattering amplitude are guaranteed by the Sturm-Liouville structure of the spin-2 mode system \cite{Chivukula:2019zkt}.  Therefore, these various ${\mathcal O}(\epsilon^2)$ corrections must all be connected in such a way that the sum rules remain satisfied. Verification of the $\mathcal{O}(s^{5})$ and $\mathcal{O}(s^{4})$ sum rules can, in this sense, be considered a cross-check of our perturbative computation.

At $\mathcal{O}(s^{3})$, the spin-0 KK modes become directly relevant. Consider the $\mathcal{O}(s^{3})$ sum rule, Eq. \eqref{eq:sr3}, in combination with the results of Table \ref{table:coupling-combinations}. At leading order, only the radion ($i=0$) contributes, and the couplings are the same as those in the five-dimensional orbifolded torus model \cite{Chivukula:2019rij}. At ${\mathcal O}(\epsilon^2)$, corrections to both the radion ($i=0)$ and GW scalar $(i>0)$ couplings are important. The $\mathcal{O}(s^{3})$ sum rule ultimately remains satisfied because these ${\mathcal O}(\epsilon^2)$ scalar corrections exactly cancel the previously-mentioned mass and wavefunction corrections within the spin-2 KK sector.

Finally, both the radion mass-squared $\mu^2_{(0)}$ and GW scalar masses-squared $\mu^2_{(i)}$ (for $i>0$) contribute directly to the $\mathcal{O}(s^{2})$ sum rule, Eq. \eqref{eq:sr4}. As was the case at $\mathcal{O}(s^{3})$, the ${\mathcal O}(\epsilon^2)$ scalar contributions exactly balance the $\mathcal{O}(\epsilon^{2})$ coupling and mass corrections of the spin-2 KK tower, thereby ensuring the cancellations described by Eq. \eqref{eq:sr4} remain satisfied despite new nontrivial contributions originating from the stabilized geometry.

Therefore, we have directly demonstrated that any contributions to the $n,n\rightarrow n,n$ scattering amplitude of helicity-0, KK level-$n$, massive spin-2 KK bosons which grow faster than $\mathcal{O}(s)$ are cancelled in the flat stabilized model at $\mathcal{O}(\epsilon^{2})$. By applying all of the sum rules previously derived, we obtain
\begin{align}
    \mathcal{M}^{(1)} &= \dfrac{\kappa^{2} s}{2048 n^{2} \pi r_{c}} \bigg\{24 n^{2} +\bigg[-69 + 16 n^{2}\pi^{2} \bigg]\epsilon^{2}\bigg\} \dfrac{(7+c_{2\theta})^{2}}{s_{\theta}^{2}} + \mathcal{O}(\epsilon^{3})~
\end{align}
as the leading high-energy scattering amplitude in the flat stabilized model at $\mathcal{O}(\epsilon^{2})$.
\section{Conclusions}

\label{sec:Conclusions}

In this paper, we calculated the scattering amplitude for tree-level elastic scattering of helicity-0, level-$n$, massive spin-2 Kaluza-Klein states within models implementing a dynamical mechanism for stabilizing the size of an extra dimension, in which the radion mode is massive. The cancellations occur only when one includes not only the massive radion, but also the massive spin-0 modes arising from the Goldberger-Wise bulk scalar introduced to stabilize the size of the extra dimension. We have derived the sum rules between the masses and couplings of the spin-2 and spin-0 KK modes of the model which must be satisfied such that the overall elastic scattering amplitude grows like $E^{2}$, despite individual contributions to the amplitude growing as fast as $E^{10}$. We introduced a simple model of a stabilized extra dimension which is a small deformation of a flat (toroidal) five-dimensional model, and we directly confirmed cancellations in the high-energy scattering amplitude to leading nontrivial order in the deformation. These results extend previous work \cite{Chivukula:2019rij,Chivukula:2019zkt,Chivukula:2020hvi,Foren:2020egq} which demonstrated related cancellations in unstabilized models.

In future work \cite{Chivukula:2020tbd}, we will present technical details regarding the interplay of components of the spin-0 sector, including how the bulk scalar and scalar metric sectors combine into a single scalar sector, and attending to contributions from the boundaries (building on the work of \cite{Kofman:2004tk,Bonifacio:2019ioc,George:2011sw}). We will also examine the scattering of massive spin-2 states in the phenomenologically motivated case of perturbing around a warped background \cite{DeWolfe:1999cp} and  discuss the phenomenological implications, in particular for the thermal history of the early universe and models involving a spin-2 dark matter portal.

\medskip

\noindent {\bf Acknowledgements:} This material is based upon work supported by the National Science Foundation under Grant No. PHY- 1915147 .
\appendix
\section[Perturbative Analysis of the
Flat Stabilized Model]{Perturbative Analysis of the Flat Stabilized Model}

\label{sec:appendix}

In this appendix, we describe the perturbative calculations leading to the wavefunctions and eigenvalues listed in Table \ref{table:wavefunctions} for the flat stabilized model. In particular, while the analysis of the spin-2 sector proceeds along the familiar lines of Rayleigh-Schr\"odinger perturbation theory, the analysis of the spin-0 sector is complicated by the $(\partial_\varphi \phi_0)^2 = \mathcal{O}(\epsilon^{2})$ quantity in the denominator of both forms of the spin-0 normalization condition, Eqs. \eqref{eq:scalarnorm1} and \eqref{eq:scalarnorm2}.

\subsection{Spin-2}

In combination with the boundary conditions $(\partial_{\varphi}\psi_{n})=0$ at $\varphi \in\{0,\pi\}$, the spin-2 mode equation, Eq. \eqref{eq:spin2-SLeqn}, defines a Sturm-Liouville problem. To evaluate this equation through $\mathcal{O}(\epsilon^{2})$ in the flat stabilized model, we note the warp factor equals $A(\varphi)= \epsilon^{2}|\varphi|^{2} + \mathcal{O}(\epsilon^{3})$. Expanding in powers of $\epsilon$, the mode equation becomes, in the bulk,
\begin{eqnarray}
    \partial_{\varphi}\bigg[ \bigg(1- 4\varphi^{2}\epsilon^{2} + \mathcal{O}(\epsilon^{3})\bigg) (\partial_{\varphi}\psi_{n})\bigg] &=& -\mu_{n}^{2} \bigg(1- 2\varphi^{2}\epsilon^{2} + \mathcal{O}(\epsilon^{3})\bigg)\psi_{n}~,
    \label{sl2flat}
\end{eqnarray}
subject to the the orthonormality condition, Eq. \eqref{eq:spin2-mode-norm},
\begin{equation}
    \frac{1}{\pi} 
    \int_{-\pi}^{+\pi} d\varphi\hspace{5 pt} \bigg(1-2|\varphi|^2\epsilon^2+\mathcal{O}(\epsilon^{3})\bigg)\, \psi_m \psi_n  = \delta_{m,n}~,
    \label{eq:sl2norm}
\end{equation}
and the boundary conditions $(\partial_{\varphi} \psi_{n}) = 0$ at $\varphi \in \{0,\pi\}$.

To evaluate this in perturbation theory, we expand the spin-2 eigenfunctions and eigenvalues in powers of $\epsilon$. Because there is no $\mathcal{O}(\epsilon)$ term in Eqs. \eqref{sl2flat} and \eqref{eq:sl2norm}, we need not include such a term in these expansions:
\begin{eqnarray}
    \psi_{n} &= &\psi_{n,0} + \epsilon^{2}\,\psi_{n,2} + \mathcal{O}(\epsilon^{3}), \\
    \mu^2_{n} &=& \mu^2_{n,0} + \epsilon^{2}\,\mu^2_{n,2} + \mathcal{O}(\epsilon^{3}).
\end{eqnarray}
At leading order, i.e. $\mathcal{O}(\epsilon^{0})$, the spin-2 mode equation equals
\begin{equation}
    \Big[\partial^{2}_{\varphi} + \mu_{n,0}^{2}\Big] \psi_{n,0} = 0~.
\end{equation}
Imposing the normalization and boundary conditions to leading order, we obtain the usual graviton and KK-modes in the five-dimensional orbifolded torus model, as shown in the second column of Table \ref{table:wavefunctions}.
 
At order $\epsilon^{2}$, Eq. \eqref{sl2flat} implies
\begin{equation}
    \Big[\partial_{\varphi}^{2} + \mu_{n,2}^{2}\Big]\psi_{n,2} = \Big[4\varphi^{2} \,\partial_{\varphi}^{2} + 8\varphi\, \partial_{\varphi} + \Big(2 \varphi^{2}\mu_{n,0}^{2}  - \mu_{n,2}^{2}\Big)\Big]\psi_{n,0}
\end{equation}
Using the previously obtained eigenfunctions and eigenvalues, and then imposing the Neumann boundary conditions $(\partial_{\varphi}\psi_{n})=0$ at $\varphi\in\{0,\pi\}$ and the normalization condition Eq. \eqref{eq:sl2norm} at $\mathcal{O}(\epsilon^{2})$, we obtain the spin-2 corrections listed in the last column of Table \ref{table:wavefunctions}.

\subsection{Spin-0}

Next, we analyze the spin-0 mode equation defined in Eq. \eqref{eq:spin0mode-equation}, subject to the normalization conditions in Eqs. \eqref{eq:scalarnorm1} and \eqref{eq:scalarnorm2}. By applying the stiff-wall limit described in Eq. \eqref{eq:stiff-wall}, these equations yield a Sturm-Liouville problem with boundary conditions $(\partial_{\varphi}\gamma_{n})=0$ at $\varphi\in\{0,\pi\}$. To perturbatively solve these equations, we need -- in addition to the warp factor $A(\varphi)=\epsilon^2|\varphi|^{2} + \mathcal{O}(\epsilon^{3})$ -- the scalar background $\phi_{0}$ defined in Eq. \eqref{eq:scalarbackground}. For our present purposes, we rewrite $\phi_{0}$ in terms of a convenient parameter $\alpha = 2\sqrt{6}/\phi_1$, such that
\begin{equation}
    \phi_0(\varphi)=\phi_{1} \, e^{-\alpha \epsilon |\varphi|}\hspace{15 pt} \Rightarrow\hspace{15 pt} (\partial_\varphi \phi_0)^2 = 24\, \epsilon^{2}\, e^{-2\alpha \epsilon |\varphi|}
\end{equation}

The presence of $(\partial_{\varphi}\phi_{0})^{2}$ in multiple denominators thus yields several factors of $1/24\epsilon^{2}$, which are inconvenient for our perturbative analysis; to eliminate these, we multiply the spin-0 mode equation by $24 \epsilon^{2}$. Having done so, expanding the spin-0 mode equation
Eq. \eqref{eq:spin0mode-equation} through $\mathcal{O}(\epsilon^{2})$ yields
\begin{eqnarray}
   &\partial_\varphi\bigg[\bigg(1  + 2\alpha\varphi\epsilon + 2 (1+\alpha^2)\varphi^{2}\epsilon^{2} + \mathcal{O}(\epsilon^{3}) \bigg)(\partial_{\varphi} \gamma_{n})  \bigg] - \bigg(4\epsilon^{2}+\mathcal{O}(\epsilon^{3})\bigg)  \gamma_{n} \label{sl0flat}\\ 
    &\hspace{35 pt}=-\mu_{(n)}^{2} \bigg(1 + 2\alpha\varphi\epsilon + 2(2+\alpha^2)\varphi^{2}\epsilon^{2} + \mathcal{O}(\epsilon^{3}) \bigg)\gamma_{n}~. \nonumber
\end{eqnarray}
in the bulk. We next solve for $\gamma_{n}$ and $\mu_{(n)}^{2}$ perturbatively in $\epsilon$; however, we consider the radion and GW scalars separately because of their different behaviors with respect to $\epsilon$.

First, we consider the GW scalars, which have eigenfunctions $\gamma_{n}$ and eigenvalues $-\mu_{(n)}^{2}$ with $n>0$. We normalize the GW scalar wavefunctions according to the form of the orthonormality condition as defined in Eq. \eqref{eq:scalarnorm2}, which already implicitly uses the stiff-wall limit. Once we have expanded in $\epsilon$ and both sides are multiplied by $24\epsilon^{2}$ as before, this yields:
\begin{equation}
    \dfrac{12}{\pi}\,\mu_{(n)}^{2}\,\int_{0}^{\pi} d\varphi\hspace{5 pt} \bigg(1 + 2\alpha\varphi\epsilon + 2(2+\alpha^2)\varphi^{2}\epsilon^{2} + \mathcal{O}(\epsilon^{3}) \bigg)\, \gamma_{m}\gamma_{n} =  24 \epsilon^{2} \, \delta_{m,n}
\label{eq:scalarnorm2-pt}
\end{equation}
where $\gamma_{n}$ is subject to the Neumann boundary conditions $(\partial_{\varphi}\gamma_{n})=0$ at $\varphi\in\{0,\pi\}$. Note the orbifold-symmetric integral has been rewritten to be over $\varphi\in(0,\pi)$ as to enable the replacement $|\varphi|\rightarrow \varphi$.

We expand the spin-0 eigenfunctions and eigenvalues perturbatively in $\epsilon$ as
\begin{eqnarray}
    \gamma_{n}(\varphi) &=& \gamma_{n,0} + \epsilon\,\gamma_{n,1}(\varphi) + \epsilon^{2}\,\gamma_{n,2}(\varphi) + \mathcal{O}(\epsilon^{3})~, \label{eq:s0eigenfunctionexpansion} \\
    \mu_{(n)}^{2} &=& \mu_{(n,0)}^{2} + \epsilon\, \mu_{(n,1)}^{2}+ \epsilon^{2}\,\mu_{(n,2)}^{2} + \mathcal{O}(\epsilon^{3})~,\label{eq:s0eigenvalueexpansion}
\end{eqnarray}
The normalization condition Eq. \eqref{eq:scalarnorm2-pt} reveals that so long as $\mu^2_{(n,0)}$ is nonzero the wavefunctions $\gamma_{n}$ must start at ${\mathcal O}(\epsilon)$. Therefore, because the masses of the GW scalars are nonzero in the unstabilized limit, they satisfy $\gamma_{n,0}\equiv 0$, and the first nonzero contribution to the spin-0 mode equation Eq. \eqref{sl0flat} occurs at $\mathcal{O}(\epsilon)$:
\begin{equation}
    \Big[\partial^{2}_{\varphi} + \mu_{(n,0)}^{2}\Big] \gamma_{n,1} = 0~.
\end{equation}
Imposing the boundary conditions and the normalization of Eq. \eqref{eq:scalarnorm2-pt} to leading nontrivial order, we calculate the GW mode contributions listed in the third column of Table \ref{table:wavefunctions}. As discussed in the text, due to the discrete momentum conservation of the unperturbed toroidal background, this computation of the GW scalar wavefunctions and masses to ${\mathcal O}(\epsilon)$ is sufficient for the verification of the sum rules discussed in this paper. We have also computed the next order corrections, however, and they are listed in the fourth column of Table \ref{table:wavefunctions} for completeness.

Lastly, we consider the radion. Here, the situation reverses that of the GW scalars: in the $\epsilon \to 0$ limit, the stabilized RS model maps onto the unstabilized five-dimensional orbifolded torus model discussed in \cite{Chivukula:2019rij}, wherein it is the radion mass that vanishes and the wavefunction that becomes a nonzero constant. Hence, a perturbative expansion of the radion eigenfunction and eigenvalue similar to those defined in Eqs. \eqref{eq:s0eigenfunctionexpansion}-\eqref{eq:s0eigenvalueexpansion} begins with $\mu^2_{(0,0)}=0$ and $\gamma_{0,0} = C \neq 0$ -- which is consistent with the leading order perturbative equation
\begin{equation}
    \Big[\partial^{2}_{\varphi} + \mu_{(0,0)}^{2}\Big] \gamma_{0,0} = 0~.
\end{equation}
defined by the spin-0 mode equation, Eq. \eqref{sl0flat}, at $\mathcal{O}(\epsilon^{0})$. At next order, using the fact that $\gamma_{0,0}$ is a constant, the $\mathcal{O}(\epsilon)$ contributions to the spin-0 mode equation yield
\begin{equation}
    \partial^2_\varphi \gamma_{0,1}=-\mu^2_{(0,1)} \gamma_{0,0}~.
\end{equation}
The right-hand side of this equation is a constant, so $\gamma_{0,1}$ could in principle be a quadratic function of $\varphi$. Imposing the Neumann boundary conditions $(\partial_{\varphi}\gamma_{n})=0$ at $\varphi\in\{0,\pi\}$, however, we find that $\gamma_{0,1}$ must also be a constant, and thus $\mu^2_{(0,1)}\equiv 0$. Finally, at $\mathcal{O}(\epsilon^{2})$, the spin-0 mode equation yields
\begin{equation}
     \partial^2_\varphi \gamma_{0,2}=\Big[4-\mu^2_{(0,2)} \Big]\gamma_{0,0}~.
\end{equation}
The same reasoning that led us through the $\mathcal{O}(\epsilon)$ computation applies once again here. Thus, after imposing the boundary conditions, we conclude $\gamma_{0,2}$ must be a constant; however, now we find a nonzero eigenvalue contribution: $\mu^2_{(0,2)}=4$ \cite{Csaki:2000zn}. In this way, the radion has become massive at $\mathcal{O}(\epsilon^{2})$.

Because $\mu^2_{(0)}={\mathcal O}(\epsilon^2)$, the radion wavefunction must be normalized via the orthonormality condition as written in Eq. \eqref{eq:scalarnorm1}. That is, we apply the orthonormality condition
\begin{eqnarray}
 \frac{12}{\pi} \int_{0}^{+\pi}d\varphi  \, \bigg[ 
 \bigg(1  + 2\alpha\varphi\epsilon + 2 (1+\alpha^2)\varphi^{2}\epsilon^{2} + \mathcal{O}(\epsilon^{3})\bigg) \dfrac{(\partial_\varphi \gamma_m) (\partial_\varphi \gamma_n)}{24 \epsilon^2}\label{eq:norm-for-radion} \\
  + \frac{1}{6}\bigg(1+2\varphi^2\epsilon^2+\mathcal{O}(\epsilon^{3})\bigg)\gamma_m \gamma_n\bigg] = \delta_{m,n}~, \nonumber
\end{eqnarray}
Note the orbifold-symmetric integral has been rewritten to be over $\varphi\in(0,\pi)$ so as to enable the replacement $|\varphi|\rightarrow \varphi$. For the radion, the first term in the integrand begins at ${\mathcal O}(\epsilon^{4})$ since the radion wavefunction is a constant through ${\mathcal O}(\epsilon^2)$; in other words, each $(\partial_{\varphi}\gamma_{n})$ is $\mathcal{O}(\epsilon^{3})$ at largest. Keeping this in mind, we use Eq. \eqref{eq:norm-for-radion} to normalize the radion wavefunction and thereby obtain the radion results shown in Table \ref{table:wavefunctions}.

\bibliographystyle{unsrt}
\bibliography{references}

\begin{thebibliography}{10}

\bibitem{Kaluza:1921tu}
Th. Kaluza.
\newblock {Zum Unit\"atsproblem der Physik}.
\newblock {\em Int. J. Mod. Phys. D}, 27(14):1870001, 2018.

\bibitem{Klein:1926tv}
Oskar Klein.
\newblock {Quantum Theory and Five-Dimensional Theory of Relativity. (In German
  and English)}.
\newblock {\em Z. Phys.}, 37:895--906, 1926.

\bibitem{Appelquist:1987nr}
T.~Appelquist, A.~Chodos, and P.~G.~O. Freund, editors.
\newblock {\em {MODERN KALUZA-KLEIN THEORIES}}.
\newblock 1987.

\bibitem{Antoniadis:1990ew}
Ignatios Antoniadis.
\newblock {A Possible new dimension at a few TeV}.
\newblock {\em Phys. Lett. B}, 246:377--384, 1990.

\bibitem{ArkaniHamed:1998rs}
Nima Arkani-Hamed, Savas Dimopoulos, and G.~R. Dvali.
\newblock {The Hierarchy problem and new dimensions at a millimeter}.
\newblock {\em Phys. Lett. B}, 429:263--272, 1998.

\bibitem{Appelquist:2000nn}
Thomas Appelquist, Hsin-Chia Cheng, and Bogdan~A. Dobrescu.
\newblock {Bounds on universal extra dimensions}.
\newblock {\em Phys. Rev. D}, 64:035002, 2001.

\bibitem{Randall:1999ee}
Lisa Randall and Raman Sundrum.
\newblock {A Large mass hierarchy from a small extra dimension}.
\newblock {\em Phys. Rev. Lett.}, 83:3370--3373, 1999.

\bibitem{Randall:1999vf}
Lisa Randall and Raman Sundrum.
\newblock {An Alternative to compactification}.
\newblock {\em Phys. Rev. Lett.}, 83:4690--4693, 1999.

\bibitem{Chivukula:2019rij}
R.~Sekhar~Chivukula, Dennis Foren, Kirtimaan~A. Mohan, Dipan Sengupta, and
  Elizabeth~H. Simmons.
\newblock {Scattering amplitudes of massive spin-2 Kaluza-Klein states grow
  only as ${\cal O}(s)$}.
\newblock {\em Phys. Rev. D}, 101(5):055013, 2020.

\bibitem{Chivukula:2019zkt}
R.~Sekhar~Chivukula, Dennis Foren, Kirtimaan~A. Mohan, Dipan Sengupta, and
  Elizabeth~H. Simmons.
\newblock {Sum Rules for Massive Spin-2 Kaluza-Klein Elastic Scattering
  Amplitudes}.
\newblock {\em Phys. Rev. D}, 100(11):115033, 2019.

\bibitem{Chivukula:2020hvi}
R.~Sekhar Chivukula, Dennis Foren, Kirtimaan~A. Mohan, Dipan Sengupta, and
  Elizabeth~H. Simmons.
\newblock {Massive Spin-2 Scattering Amplitudes in Extra-Dimensional Theories}.
\newblock {\em Phys. Rev. D}, 101(7):075013, 2020.

\bibitem{Foren:2020egq}
Dennis Foren.
\newblock {\em {Scattering Amplitudes in Theories of Compactified Gravity}}.
\newblock PhD thesis, Michigan State U., 2020.

\bibitem{Bonifacio:2019ioc}
James Bonifacio and Kurt Hinterbichler.
\newblock {Unitarization from Geometry}.
\newblock {\em JHEP}, 12:165, 2019.

\bibitem{ArkaniHamed:2002sp}
Nima Arkani-Hamed, Howard Georgi, and Matthew~D. Schwartz.
\newblock {Effective field theory for massive gravitons and gravity in theory
  space}.
\newblock {\em Annals Phys.}, 305:96--118, 2003.

\bibitem{ArkaniHamed:2003vb}
Nima Arkani-Hamed and Matthew~D. Schwartz.
\newblock {Discrete gravitational dimensions}.
\newblock {\em Phys. Rev. D}, 69:104001, 2004.

\bibitem{Hinterbichler:2011tt}
Kurt Hinterbichler.
\newblock {Theoretical Aspects of Massive Gravity}.
\newblock {\em Rev. Mod. Phys.}, 84:671--710, 2012.

\bibitem{Rattazzi:2003ea}
R.~Rattazzi.
\newblock {Cargese lectures on extra-dimensions}.
\newblock In {\em {Cargese School of Particle Physics and Cosmology: the
  Interface}}, pages 461--517, 8 2003.

\bibitem{Maldacena:1997re}
Juan~Martin Maldacena.
\newblock {The Large N limit of superconformal field theories and
  supergravity}.
\newblock {\em Int. J. Theor. Phys.}, 38:1113--1133, 1999.

\bibitem{Witten:1998qj}
Edward Witten.
\newblock {Anti-de Sitter space and holography}.
\newblock {\em Adv. Theor. Math. Phys.}, 2:253--291, 1998.

\bibitem{Aharony:1999ti}
Ofer Aharony, Steven~S. Gubser, Juan~Martin Maldacena, Hirosi Ooguri, and Yaron
  Oz.
\newblock {Large N field theories, string theory and gravity}.
\newblock {\em Phys. Rept.}, 323:183--386, 2000.

\bibitem{Brans:1961sx}
C.~Brans and R.~H. Dicke.
\newblock {Mach's principle and a relativistic theory of gravitation}.
\newblock {\em Phys. Rev.}, 124:925--935, 1961.

\bibitem{Abu-Ajamieh:2017khi}
Fayez Abu-Ajamieh, Jun~Seok Lee, and John Terning.
\newblock {The Light Radion Window}.
\newblock {\em JHEP}, 10:050, 2018.

\bibitem{Schwartz:2003vj}
Matthew~D. Schwartz.
\newblock {Constructing gravitational dimensions}.
\newblock {\em Phys. Rev. D}, 68:024029, 2003.

\bibitem{Goldberger:1999uk}
Walter~D. Goldberger and Mark~B. Wise.
\newblock {Modulus stabilization with bulk fields}.
\newblock {\em Phys. Rev. Lett.}, 83:4922--4925, 1999.

\bibitem{Goldberger:1999un}
Walter~D. Goldberger and Mark~B. Wise.
\newblock {Phenomenology of a stabilized modulus}.
\newblock {\em Phys. Lett. B}, 475:275--279, 2000.

\bibitem{Tanaka:2000er}
Takahiro Tanaka and Xavier Montes.
\newblock {Gravity in the brane world for two-branes model with stabilized
  modulus}.
\newblock {\em Nucl. Phys. B}, 582:259--276, 2000.

\bibitem{Csaki:2000zn}
Csaba Csaki, Michael~L. Graesser, and Graham~D. Kribs.
\newblock {Radion dynamics and electroweak physics}.
\newblock {\em Phys. Rev. D}, 63:065002, 2001.

\bibitem{Kofman:2004tk}
Lev Kofman, Johannes Martin, and Marco Peloso.
\newblock {Exact identification of the radion and its coupling to the
  observable sector}.
\newblock {\em Phys. Rev. D}, 70:085015, 2004.

\bibitem{Boos:2005dc}
Edward~E. Boos, Yuri~S. Mikhailov, Mikhail~N. Smolyakov, and Igor~P. Volobuev.
\newblock {Physical degrees of freedom in stabilized brane world models}.
\newblock {\em Mod. Phys. Lett. A}, 21:1431--1449, 2006.

\bibitem{George:2011sw}
Damien~P. George and Kristian~L. McDonald.
\newblock {Gravity on a Little Warped Space}.
\newblock {\em Phys. Rev. D}, 84:064007, 2011.

\bibitem{Chivukula:2020tbd}
R.~Sekhar Chivukula, Dennis Foren, Kirtimaan~A. Mohan, Dipan Sengupta, and
  Elizabeth~H. Simmons.
\newblock {work in progress}.
\newblock 2021.

\bibitem{Charmousis:1999rg}
Christos Charmousis, Ruth Gregory, and V.~A. Rubakov.
\newblock {Wave function of the radion in a brane world}.
\newblock {\em Phys. Rev. D}, 62:067505, 2000.

\bibitem{DeWolfe:1999cp}
O.~DeWolfe, D.~Z. Freedman, S.~S. Gubser, and A.~Karch.
\newblock {Modeling the fifth-dimension with scalars and gravity}.
\newblock {\em Phys. Rev. D}, 62:046008, 2000.

\end{thebibliography}

\end{document}